\documentclass[aps,prl,superscriptaddress,twocolumn,showpacs]{revtex4-1}

\usepackage{amsmath, amsthm, amssymb}
\usepackage{graphicx}
\usepackage{subfigure}
\usepackage{color}
\usepackage{times}
\usepackage{float}
\usepackage{natbib}
\usepackage{hyperref}

\usepackage{soul,xcolor}

\graphicspath{{./}{../Figures/}}

\hypersetup{pdfpagemode=UseNone}

\hypersetup{
        colorlinks=true,
}

\newcommand{\beq}    {\begin{equation}}
\newcommand{\enq}    {\end{equation}}
\newcommand{\ceq}[1] {(\ref{#1})}

\newcommand{\pp}{{\bf p}}
\newcommand{\df}     {\equiv}

\newcommand{\rr}     {{\bf r}}
\newcommand{\RR}     {{\bf R}}
\newcommand{\sss}     {{\bf S}}

\newcommand{\sbbb}{{\bf S}}

\DeclareMathOperator{\sgn}{sgn}

\newcommand{\be}{\begin{equation} }
\newcommand{\ee}{\end{equation} }
\newcommand{\ba}{\begin{eqnarray} }
\newcommand{\ea}{\end{eqnarray} }

\newcommand{\up}{\uparrow}
\newcommand{\dn}{\downarrow}
\renewcommand{\vec}[1]{\mathbf{ #1 }}

\DeclareMathOperator{\re}{Re}
\DeclareMathOperator{\im}{Im}

\begin{document}
\setstcolor{red}

\title{Impurity-induced bound states in superconductors with spin-orbit coupling}

\author{Younghyun Kim}
\affiliation{Department of Physics, University of California,  Santa Barbara, California 93106, USA}

\author{Junhua Zhang}
\affiliation{Department of Physics, College of William and Mary, Williamsburg, Virginia 23187, USA}

\author{E. Rossi}
\affiliation{Department of Physics, College of William and Mary, Williamsburg, Virginia 23187, USA}

\author{Roman M. Lutchyn}
\affiliation{Station Q, Microsoft Research, Santa Barbara, California 93106-6105, USA}

\date{\today}

\begin{abstract}
We study the effect of strong spin-orbit coupling (SOC) on bound states
induced by impurities in superconductors.
%
%
The presence of spin-orbit coupling breaks the $\mathbb{SU}(2)$-spin symmetry and causes the
superconducting order parameter to have generically both singlet
(s-wave) and triplet (p-wave) components. We find that in the presence of SOC the spectrum of Yu-Shiba-Rusinov (YSR) states is qualitatively different in s-wave and p-wave superconductor, a fact that can be used to identify the superconducting pairing symmetry of the host system.
We also predict that in the presence of SOC the spectrum of the impurity-induced bound states depends on the orientation of the magnetic moment {\bf S} of the impurity and, in particular,
that by changing the orientation of {\bf S} the fermion-parity of the lowest energy bound state can be tuned.
We then study the case of a dimer of magnetic impurities and show that in this case the YSR spectrum for a p-wave
superconductor is qualitatively very different from the one for an s-wave superconductor even in the
limit of vanishing SOC. Our predictions can be used to distinguish the symmetry of the order parameter
and have implications for the Majorana proposals based on
chains of magnetic atoms placed on the surface of superconductors with strong spin-orbit coupling~\cite{nadjperge2014}.

\end{abstract}


\maketitle

The presence of impurities is almost always unavoidable in condensed matter systems.
Often impurities are regarded as a nuisance that spoils the properties
of a clean system and complicates the understanding of its properties.
However, impurities are in many instances essential to obtain desirable physical effects and
can be used as unique atomic-scale probes of the
ground state of the host system~\cite{Poilblanc'94, Flatte'97, flatte1997, Salkola97,  Yazdani'97, Hudson'99, lang'02, morr2003, balatsky2006}.
%
%
%
The study of the effect of impurities in superconductors
has been a very active field of research
\cite{balatsky2006}.
In an s-wave superconductor magnetic impurities cause the formation
of bound states,
the Yu-Shiba-Rusinov (YSR) states
\cite{yu1965,shiba1968,rusinov1969}.
There has been a significant interest in the
properties of YSR states due to theoretical proposals suggesting that a chain  of
magnetic impurities
placed on the surface of a superconductor (SC) would be a very robust, self-tuning, system
that should exhibit non-abelian, Majorana, states
\cite{nadj2013,klinovaja2013,braunecker2013,vazifeh2013, pientka2013}.
In these proposals the bound states induced by the chain of magnetic impurities
form an impurity band with non-trivial topological character.
More recently it has been pointed out that the presence
of Rashba spin-orbit coupling (SOC) should facilitate the realization of
a topological impurity band of YSR states.
\cite{kim2014, nadjperge2014, brydon2014,  Ebisu14}.
On the surface, due to the lack of inversion symmetry, some
amount of Rashba SOC will be present.
Therefore, for the systems considered to realize a topological band of YSR states
the presence of Rashba SOC is both
unavoidable and beneficial.
This assessment has very recently been confirmed
by the experimental results presented in Ref.~\onlinecite{nadjperge2014},
that show some evidence of the presence of Majorana modes at the end
of a chain of Fe atoms placed on the surface of a SC with strong SOC,  Pb.
The recent developments in the search of systems that
can resiliently host Majorana fermions~\cite{Fu&Kane08, Fu&Kane09, Sau10, Alicea10, LutchynPRL10, 1DwiresOreg, Linder10, MajoranaTInanowires, 1DwiresLutchyn2, Duckheim'11, Choy'11, SB'11, Flensberg'12, Potter'12, Martin'12, SauNature'12, Mourik2012, Rokhinson2012, Das2012, Deng2012, Fink2012, Churchill2013, hui2014, Li14} strongly motivates the
study of the effect of SOC on YSR states. However, so
far the effects of SOC on YSR states have been almost completely neglected.
%

In this work we present the general theory of the impurity-induced bound states in the presence of Rashba SOC.
We show that SOC, which breaks $\mathbb{SU}(2)$-spin symmetry and results in the mixture of  s-wave and p-wave pairing correlations~\cite{Gorkov01}
profoundly modifies the spectrum of the YSR states.
%
%
%
%
Our theory takes into account the fact that the impurity potential normally has both a scalar and a magnetic component.
We consider the realistic, and general, case in which both the scalar and the magnetic part of the impurity potential
has angular momentum components ($l$) higher than $l=0$.
This is also motivated by the fact that
partial waves beyond $s$-wave have been shown to
often be essential to explain experimental data~\cite{Kunz'80, ji2008, Grothe'12}.
We find that the presence of SOC, by mixing YSR states with different $l$,
profoundly changes the spectrum of the impurity-induced bound states.
%
The presence of SOC can lead to p-wave pairing. Our results show that
the spectrum of impurity-induced bound states  is qualitatively different
in p-wave and s-wave superconductors: we find that, in general, the parity of the particle (or hole)-like subgap bound states
in p-wave SCs is different from that of s-wave SCs.
%
This qualitative difference
can be used to identify the superconducting pairing symmetry of the host system.
Another important consequence of the presence of SOC that we find
is that the spectrum of the YSR states becomes dependent on the orientation of the magnetic moment
{\bf S} of the impurity and that in particular the fermion parity of the lowest energy bound state
can be tuned by changing the direction of {\bf S}.
%
%
We then study the case of a dimer formed by two magnetic impurities and find that in this case,
even in the limit of zero SOC the YSR spectrum is qualitatively different between s-wave SCs
and p-wave SCs.
%
Our results provide clear qualitative predictions that can be tested experimentally and that are directly relevant to recent
scanning-tunneling-spectroscopy (STS) measurements of the states induced
in thin films of Pb by the presence of magnetic adatoms
\cite{ji2008}.
By showing that the YSR spectrum can be modified by changing the orientation of {\bf S}
our results show an additional degree of tunability of the properties of the impurity-bound states in SCs
that could be extremely helpful to realize, and verify, the conditions necessary to
obtain a topological band of YSR states hosting Majorana zero-energy modes.
%
%

{\it Model.}
We consider a superconductor described by the mean-field Hamiltonian
${\cal{H}}_{\rm SC}=\sum_{\pp}\psi^\dagger_\pp H_{\rm SC}(\pp)\psi_\pp$
where
$\psi_{\pp}$ is the Nambu spinor
$(c_{\pp\up},c_{\pp\dn},c^\dagger_{-\pp\dn},-c^\dagger_{-\pp\up})^T$,
with $c^\dagger_{\pp\sigma}$ ($c_{\pp\sigma}$) the creation
(annihilation) operator for an electron with momentum $\pp=(p_x,p_y)$
and spin $\sigma$,
and
\be
H_{\rm SC}(\pp)=\tau_z\otimes(\xi_\vec{p}+\alpha \vec{l}_\vec{p}\cdot\boldsymbol\sigma)+\tau_x\otimes(\Delta_s+\frac{\Delta_t}{p_F}\vec{l}_\vec{p}\cdot\boldsymbol\sigma).
\label{eq:H_sc}
\ee
$\cal{H}_{\rm SC}$ describes effectively two-dimensional superconducting thin films, and
surfaces of 3D superconductors with strong Rashba SOC.
In \ceq{eq:H_sc} $\hbar=1$,
$\tau_j$, $\sigma_i$ are the Pauli matrices in Nambu and spin space respectively,
$\xi_\vec{p}=p^2/2m-\epsilon_F$,
with $m$ the effective mass of the fermionic quasiparticles; $\epsilon_F$ and $p_F=\sqrt{2m\epsilon_F}$ are the Fermi energy and Fermi momentum, respectively,
$\vec{l}_\vec{p}=(p_y,-p_x)$\cite{Frigeri04},
$\alpha$ is the strength of the Rashba SOC,
and $\Delta_s$, $\Delta_t$ are the singlet, triplet, pairing order parameters respectively, that, without
loss of generality, we take to be real.
%
%

In the presence of impurities the term
$H_{\rm imp} =\sum_j \hat V_j(|\rr-\RR_j|)= \sum_j \hat U(|\rr-\RR_j|)\tau_z\otimes\sigma_0 + \hat J(|\rr-\RR_j|)\tau_0\otimes\sbbb_j\cdot{\boldsymbol\sigma}$
must be added to $H_{\rm SC}$.
$\RR_i$s are the positions of the impurities, and $\hat U$ and $\hat J$ are the charge and magnetic potential respectively. Without loss of generality, we set $\RR=0$ for single impurity and $\RR_i=x_i$ for dimer.
Using the density of states (per spin)
$\nu_F= m/2\pi$, and the Fermi velocity $v_F= p_F/m$, we can define the dimensionless
potentials $U\df\hat U\pi\nu_F$, $J\df\hat J \pi\nu_F|{\bf S}|$,
and the dimensionless Rashba SOC  $\tilde{\alpha}\df\alpha/v_F$ which are used in the remainder of the paper.
%

%

To find the spectrum $\{E\}$ of the impurity-induced states we have to solve the Schr\"odinger equation $(H_{\rm SC}+H_{\rm imp})\psi(\rr)=E\psi(\rr)$. Let $G=[E-H_{\rm SC}]^{-1}$, then the Schr\"odinger equation can be rewritten as $[1-G(E,\rr)H_{\rm imp}]\psi(\rr)=0$ \cite{pientka2013}. The spectrum of the impurity bound states is obtained by finding the values of $E$ such that $\det[1-G(E,\rr)H_{\rm imp}]=0$.
In momentum space the Schr\"odinger equation takes the form:
\be
\psi(\vec{p})\!=\!\sum_j\! G(E,\vec{p})\!\int \!\! d\vec{p'}e^{ix_j(p\cos\theta-p'\!\cos\theta)} \hat V_j(|\vec{p}-\vec{p'}|)\psi(\vec{p'}).
\label{eq:bdg}
\ee
Following the formalism of Ref. \cite{Gorkov01}, the Green's function
$G$ can be written as the sum ($G(E,\vec{p})=[G^+(E,\vec{p})+G^-(E,\vec{p})]/2$) of the two spin helical bands
\be
G^{\pm}(E,\vec{p})=
\left( \begin{array}{cc}
E+\xi_\pm & \Delta_\pm \\
\Delta_\pm & E-\xi_\pm
\end{array}\right)\otimes\frac{\sigma_0 \pm \sin\theta\sigma_x \mp \cos\theta\sigma_y}{E^2-\xi^2_\pm-\Delta_\pm^2}.\label{fig:gpm}\nonumber
\ee
Here $p=|\pp|$, $\xi_\pm=p^2/2m\pm \alpha p-\epsilon_F$ and $\Delta_\pm=\Delta_s\pm \Delta_t p/p_F$.
Let us define $\overline{\psi_{j,\theta}}=\int \frac{pdp}{2\pi} e^{-ix_jp\cos\theta}\psi(\vec{p})$ and $\overline{G^{ij}(E,\theta)}=\int\frac{pdp}{2\pi} e^{-i(x_i-x_j)p\cos\theta} \,G(E,\vec{p})$.
Assuming that $\hat V(\pp)$ at the Fermi surface depends weakly on $p$ and integrating Eq.~\ceq{eq:bdg} with respect to $p$, we find
\be
\overline{\psi_i(\theta)}=\sum_j\overline{\hat G^{ij}(E,\theta)}\frac{1}{2\pi}\int d\theta'\hat V_j(\theta-\theta')\overline{\psi_j(\theta)}.
\label{eq:bdg2}
\ee
Rewriting all the functions of angle that enter Eq.~\ceq{eq:bdg2} in terms of their angular momentum components: $f(\theta)=\sum_l f_le^{i l\theta}$ we find:
\be
\overline{\psi_{i,l}}-\sum_{j,n} \overline{G^{ij}_n(E)}\hat V_j^{l-n}\overline{\psi_{j,l-n}}=0,
\label{eq:psim}
\ee
%
%
where
\be
\hat V_j^l=\left( \begin{array}{cc}
U_l\sigma_0 +J_l \frac{\vec{S_j}\cdot\vec{\boldsymbol\sigma}}{|\vec{S_j}|} & 0 \\
0 & -U_{-l}\sigma_0 +J_{-l}\frac{\vec{S_j}\cdot\vec{\boldsymbol\sigma}}{|\vec{S_j}|}
\end{array}\right).
\ee
Since $H_{\rm imp}$ is Hermitian and even with respect to $\theta-\theta'$, we require $U_l(=U_{-l})$ and $J_l(=J_{-l})$ to be real.
The local term $\overline{G^{ii}_n}=(\overline{G_n^+(E)}+\overline{G_n^-(E)})/2=0$ for $|n|\geq2$.
%
%
%
%
%
%
The details of the calculation are presented in the supplementary material\cite{sm}.
Henceforth, we assume that the impurity potential has only large $l=0,1$ components and neglect higher angular momentum channels.

We consider two different phases of a non-centro-symmetric SC~\cite{Sato09, Tewari'11, SauDemler'13}: s-wave ($|\Delta_s| \gg |\Delta_t|$)
and p-wave ($|\Delta_s| \ll |\Delta_t|$) pairing dominating regimes.
As we show below, the spectra are qualitatively different in the two regimes.

{\it Single magnetic impurity.}
%
The main effect  of the presence of SOC on the YSR spectrum is well exemplified by the case of purely magnetic
impurities. For this reason in the remainder we consider only purely magnetic impurities ($U_l=0$)
and discuss  in the supplementary material the case in which also a scalar component of the impurity potential is present.
%

%

\begin{figure}
	\centering
  \includegraphics[width=3.8cm]{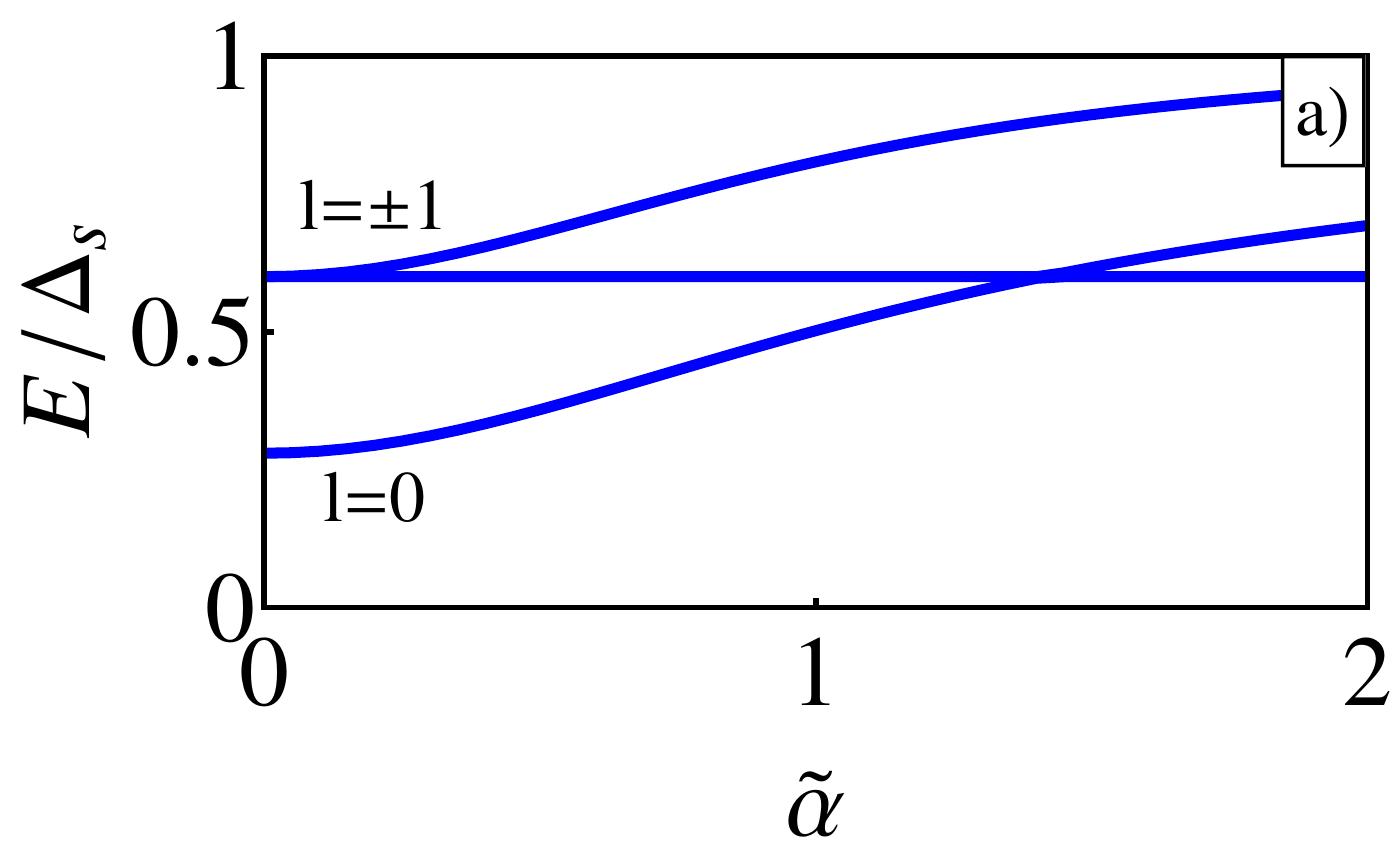} \,\,\,\,\,\,
  \includegraphics[width=3.8cm]{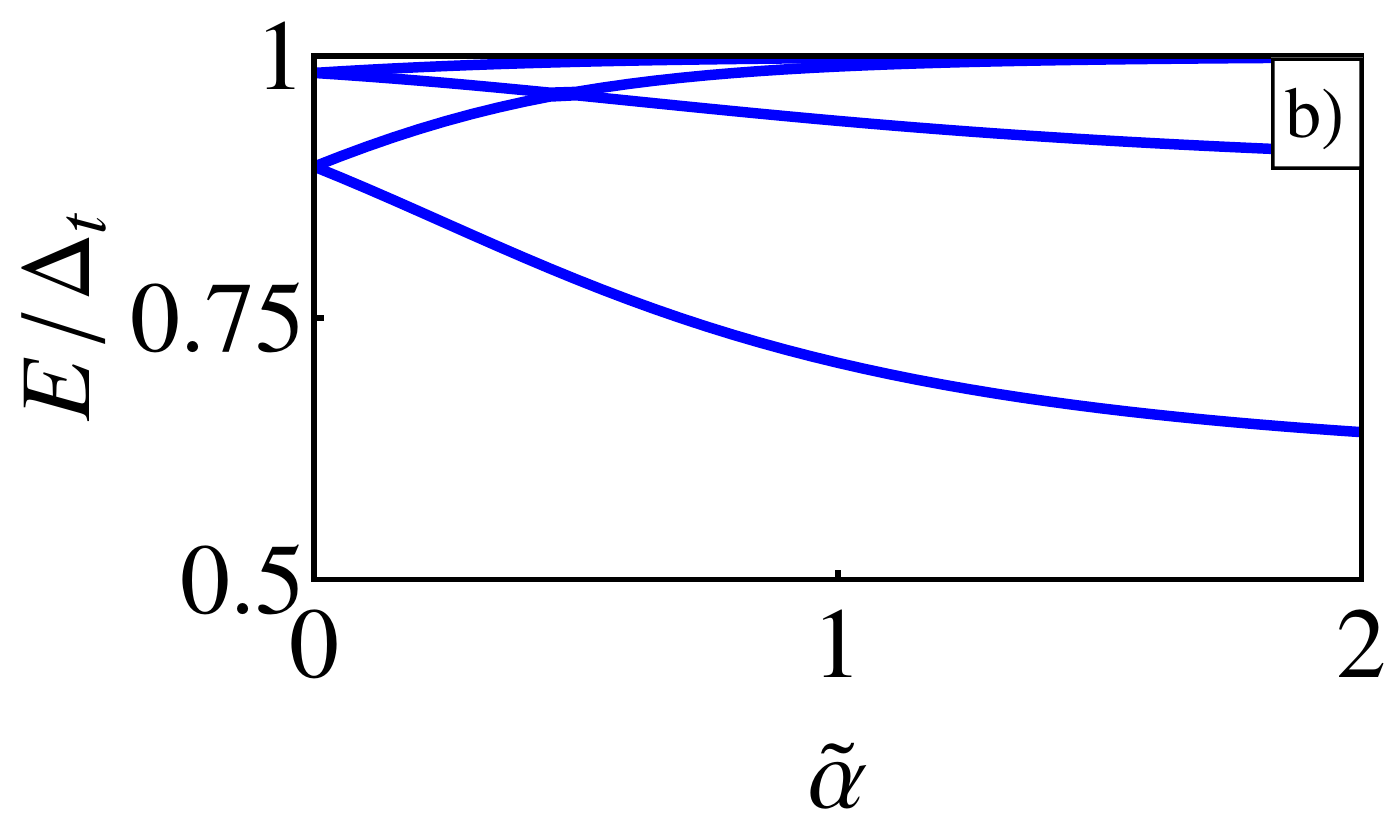} \\
  \includegraphics[width=3.8cm]{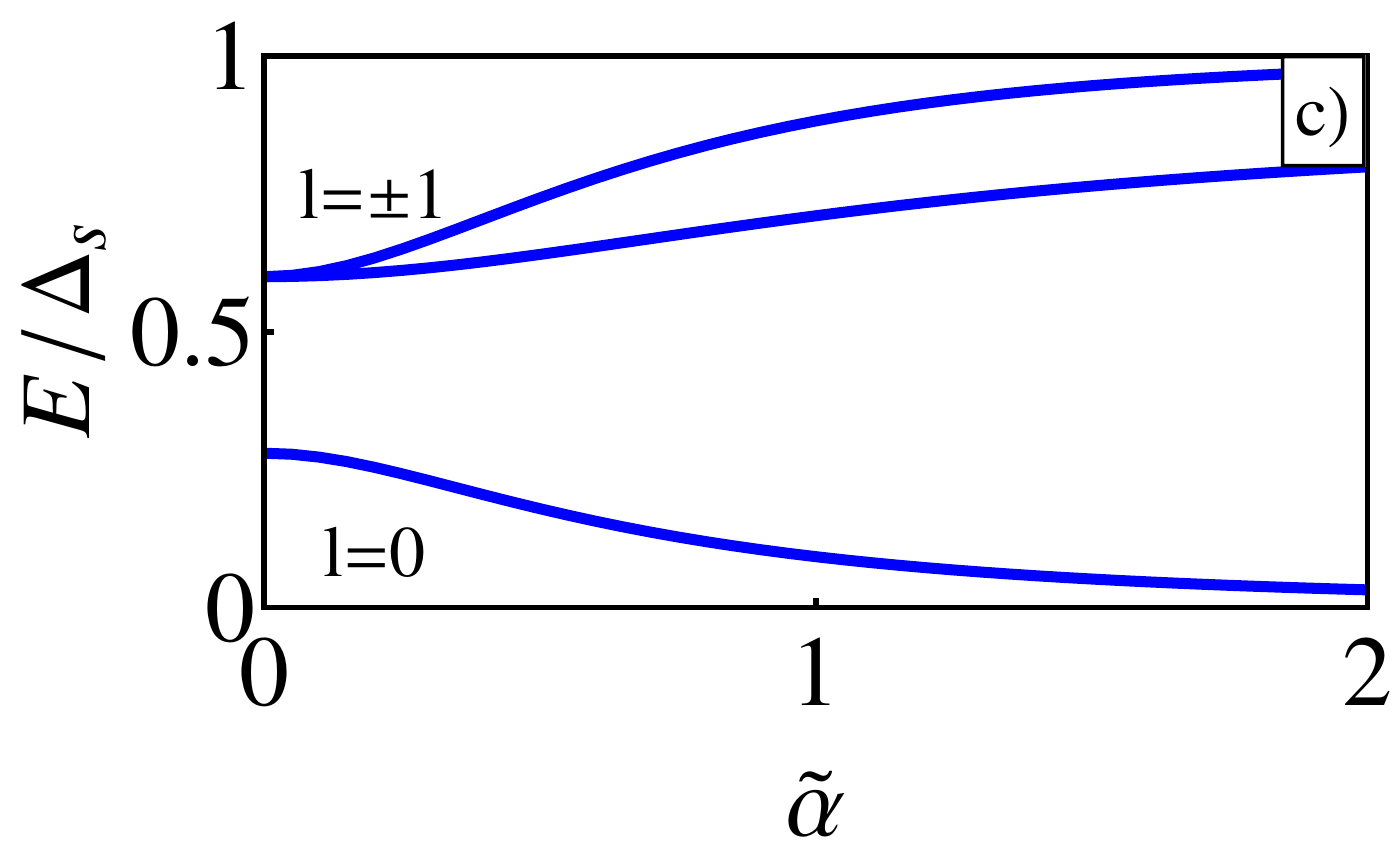}\,\,\,\,\,\,
  \includegraphics[width=3.8cm]{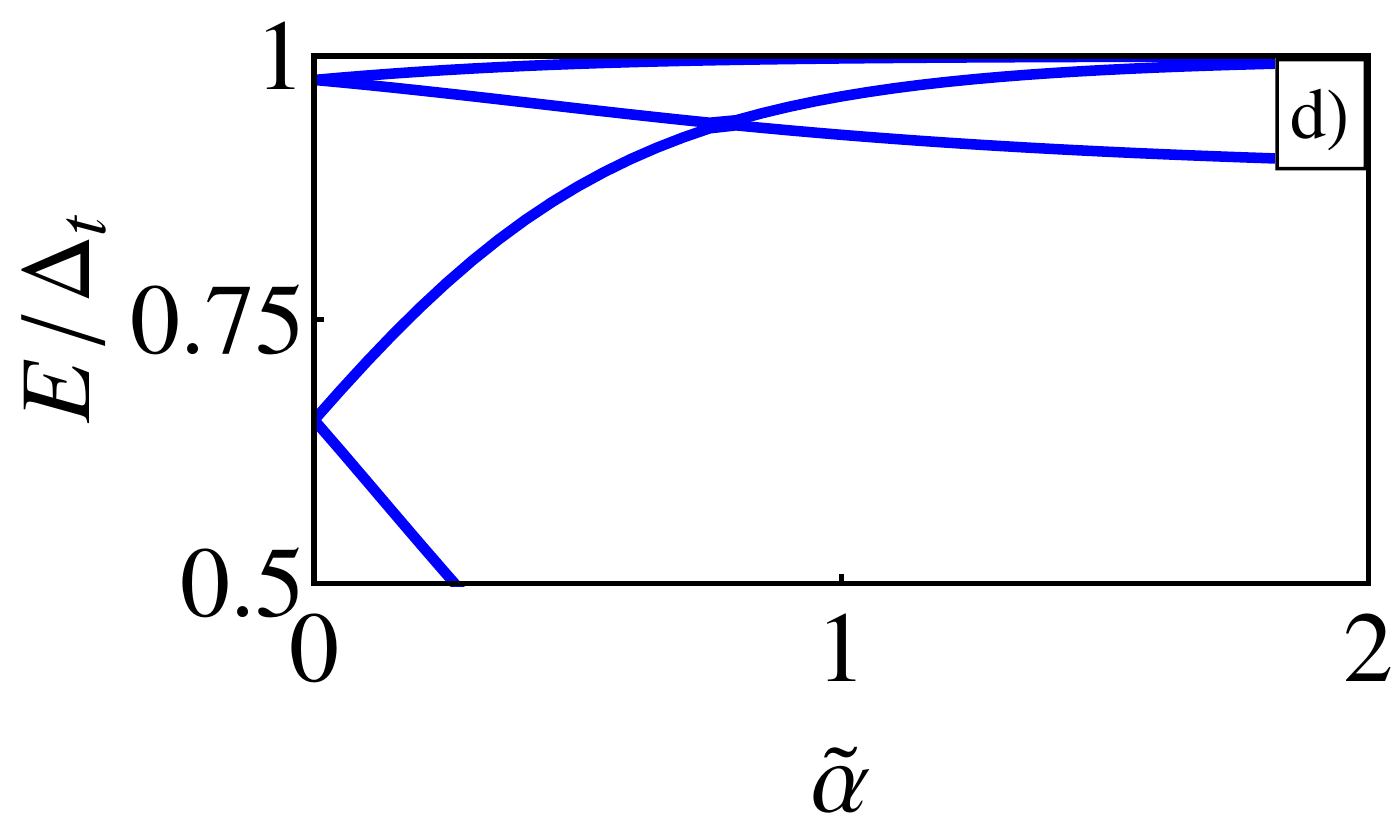} \\
\caption{
         Dependence on SOC strength of the spectrum of bound states induced in a SC by a purely magnetic impurity with $J_0 =3/4$, $J_{±1} =1/2$ in
s-wave (a, c) and p-wave (b,d) superconductor for $S\parallel \hat z$ (a,b) and $S\parallel \hat x$ (c,d).
        }
\label{fig1}
\end{figure}

For an s-wave SC, we find that, in the presence of SOC we have three impurity-induced bound states at $E>0$.
For the case when the magnetic moment of the impurity is perpendicular to the surface
of the SC, $\sss\parallel \hat z$, the energies of these states are given by the following expressions:
%
%
%
\begin{align}
\!\!\frac{|E_{1,2}|}{\Delta_s}\!&=\!\frac{\gamma^2\!-\! J_0^2J_1^2\!\pm\!\gamma^{\frac{3}{2}}\!\sqrt{(J_0^2\!-\!J_1^2)^2\!+\!(\gamma\!-\!1)(J_0\!-\!J_1)^4}}{\gamma^2(1\!+\!(J_0\!-\!J_1)^2)\!+\!2\gamma J_0J_1+J_0^2J_1^2}\\
\frac{|E_{3}|}{\Delta_s}&=\frac{1-J_1^2}{1+J_1^2}
\label{eq:e_l-1}
\end{align}
where $\gamma=1+\tilde{\alpha}^2$.
In the limit of no SC, each non-zero angular momentum component of the magnetic impurity potential, $J_n$, creates a bound state \cite{rusinov1969}.
For $\tilde{\alpha}=0$,
the $l=\pm 1$ levels are degenerate due to the rotational symmetry of the Hamiltonian.
%
%
The presence of SOC, however, causes the $l=\pm 1$ levels to split, see Fig.~\ref{fig1}~(a). Interestingly, we find that only two of the levels disperse with $\alpha$ and one level remains unchanged.
%

\begin{figure}[t]
    \includegraphics[width=3.5cm]{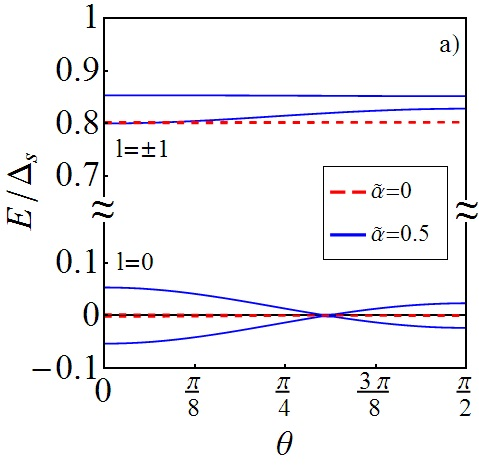}\,\,\,\,\,\,\,\,\,
    \includegraphics[width=3.3cm]{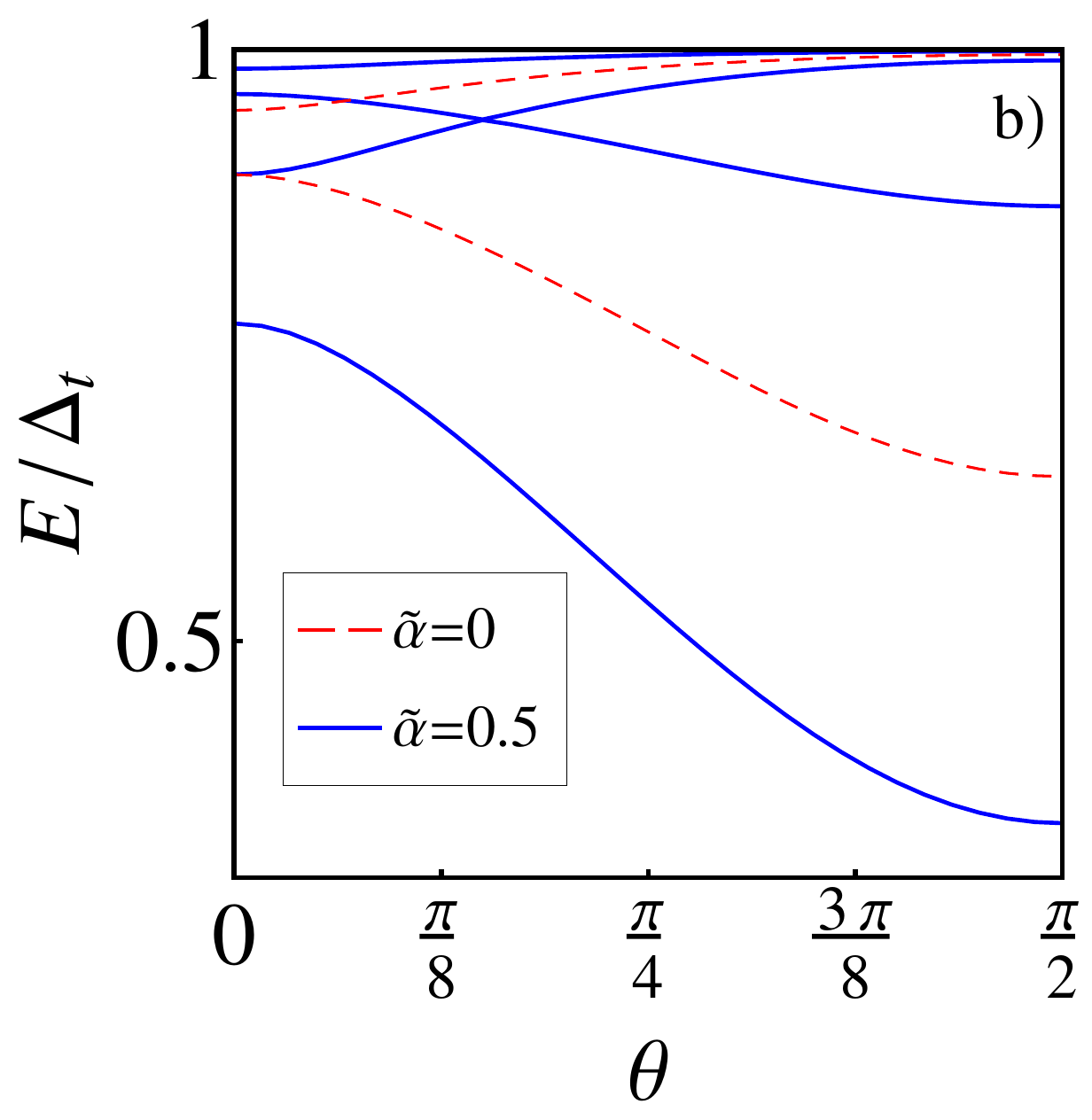}
\caption{Bound state spectrum for magnetic impurity in a s-wave(a) and p-wave(b) SC as a function of the direction of magnetic moment at $J_0=1$, $J_{\pm1}=1/3$.
}
\label{fig2}
\end{figure}

An important consequence of the presence of the SOC in s-wave SCs is that, by breaking the SU(2) symmetry of the SC Hamiltonian,
it causes the spectrum of the YSR states to strongly depend on the direction of ${\bf S}=({\cos\phi\sin\theta,\sin\phi\sin\theta,\cos\theta})$.
Fig.~\ref{fig2}~(a) shows an example of the evolution of the spectrum of the YSR states
with $\theta$ for an s-wave SC. (Due to the remaining U(1) symmetry the spectrum does not depend on the in-plane direction, i.e. $\phi$).
We see that, the spectrum for the case in which $\sss\parallel\hat z$ can be very different from the spectrum
for the case in which $\sss$ lies in the plane.
%
In particular the results of Fig.~\ref{fig2}~(a) show that by  tuning the direction of $\sss$
the fermion parity of the bound states can be changed.
This feature could be extremely useful to tune between topological and non-topological regimes in the YSR-based Majorana proposals~\cite{pientka2013}.
In the limit $\tilde{\alpha}\ll {\rm min } \{1,  |J_0-J_1|\}$ we can obtain
analytic expression for the dependence of the YSR energy levels
on the direction of $\sss$ in an s-wave SC:
%
%
%
%
\ba
\frac{|E_{1}|}{\Delta_s}&\approx&\frac{1\!-\!J_0^2}{1\!+\!J_0^2}\!+\!\frac{4\tilde{\alpha}^2J_0^2J_1(J_0\cos^2\theta-J_1)}{(1+J_0^2)^2(J_0^2-J_1^2)}\\
\frac{|E_{2,3}|}{\Delta_s}&\approx &\frac{1\!-\!J_1^2}{1\!+\!J_1^2}\!+\!\frac{2\tilde{\alpha}^2J_0J_1^2(J_0\!-\!J_1\cos^2\theta\pm F(\theta))}{(1+J_1^2)^2(J_0^2-J_1^2)}\nonumber
\ea
where $F=\sqrt{(J_0-J_1)^2\cos^2\theta+J_1^2\sin^4\theta}$.
These expressions, valid as long as the hybridized states are not degenerate,
allow us to identify the effect of the interplay of SOC, relative strength of
the different components of the magnetic impurity potentials ($J_l$), and direction
of $\sss$ on the YSR spectrum.

We now study YSR states in a p-wave SC.  The energies of the YSR spectrum,
in the presence of small SOC ($\tilde{\alpha} \ll  1$) for $\sss\parallel \hat z$ are given by
\begin{align}
\frac{|E_{1,2}|}{|\Delta_t|}&=\frac{1+J_0 J_1}{\sqrt{(1+J_0^2)(1+J_1^2)}}\pm |\tilde{\alpha}|\frac{(J_0-J_1)^2}{(1+J_0^2)(1+J_1^2)}\\
\frac{|E_{3,4}|}{|\Delta_t|}&=\frac{1}{\sqrt{1+J_1^2}}\pm |\tilde{\alpha}|\frac{J_1^2}{1+J_1^2}.
\end{align}
Fig.~\ref{fig1} show the evolution with $\tilde\alpha$ of the energies of the YSR states in a p-wave SC for $\sss\parallel \hat z$ (b) and $\sss\parallel \hat x$ (d). In the absence of SOC $\tilde{\alpha}=0$, one can see that the YSR spectrum is isotropic in s-wave case due to the rotational spin symmetry. In p-wave case, this is not the case as follows from Fig.~\ref{fig1} b) and d). Furthermore, one can notice that the states are doubly degenerate at $\tilde{\alpha}=0$ due to an additional symmetry present in the p-wave case. Indeed, the p-wave Green's function is invariant under the transformation $U=\tau_z\otimes \sigma_0 \otimes P$ with $P$ being the momentum inversion operator $\vec{p}\rightarrow-\vec{p}$. Due to this symmetry YSR states appear in pairs in p-wave superconductor. In contrast, the s-wave Green's function does not have above symmetry and, as a result, there is only one bound state per angular momentum channel (i.e one state for $l=-1,0,1$ channels). In the presence of perturbations not commuting with $U$ such as, for example, SOC, this degeneracy is lifted and the different parity of the particle (or hole)-like subgap states in s-wave and p-wave becomes visible, see Figs.~\ref{fig1} and ~\ref{fig2}.
This qualitative result opens the possibility to identify the dominant superconducting pairing of a SC
by simply counting the number of particle-like energy levels induced by a magnetic impurity within the SC gap.
%
%
%
We now discuss the dependence of the YSR spectrum on the orientation of magnetic impurity moment in p-wave superconductors. In contrast to s-wave superconductors, the YSR spectrum in p-wave case depends on $\theta$ even in the absence of SO coupling since p-wave pairing is characterized by the vector $l_{\bf p}$, see Eq.\eqref{eq:H_sc}. The analytical results for a general angle $\theta$ are not particularly illuminating (see Eq. (S26) in the supplementary information) so we plot the evolution of the YSR spectrum with $\theta$ in Fig.~\ref{fig2}~(b). One can notice that the presence of the SOC enhances the dispersion of YSR states with $\theta$.

{\it Dimer.}
There is currently a great interest in the properties of the bound states created
by a chain of magnetic impurities placed on a SC
\cite{nadj2013,klinovaja2013,braunecker2013,vazifeh2013, pientka2013, kim2014, nadjperge2014, brydon2014,  Ebisu14}.
To understand the physics of a chain of impurities it is very helpful
to investigate the simpler case of a dimer formed by two magnetic impurities.
Using Eq.~(\ref{eq:psim}) we have studied the properties of a dimer formed by two magnetic impurities placed at a distance $d$ from each other on the surface of the SC assuming $\Delta/\epsilon_F \ll 1$. We find that the wavefunction overlap between the bound states induced by the two impurities generates level splitting which strongly depends on the relative direction of the impurity spins, and that such splitting depends on the strength of the SOC. It is interesting to note that for the case of a dimer the presence of SOC, even when the SC is s-wave, modifies the spectrum also in the limit in which the magnetic part of the single impurity potential has only one nonzero angular momentum component. For this reason, to understand the effect of SOC on the YSR spectrum of a dimer we consider the case in which only $J_0$, or $J_1$ are not zero and one impurity has  $\sss_1\parallel \hat z$ and the other  $\vec{S_2}=(\sin\theta,0,\cos\theta)$.
\begin{figure}[t]
 \includegraphics[width=3.6cm]{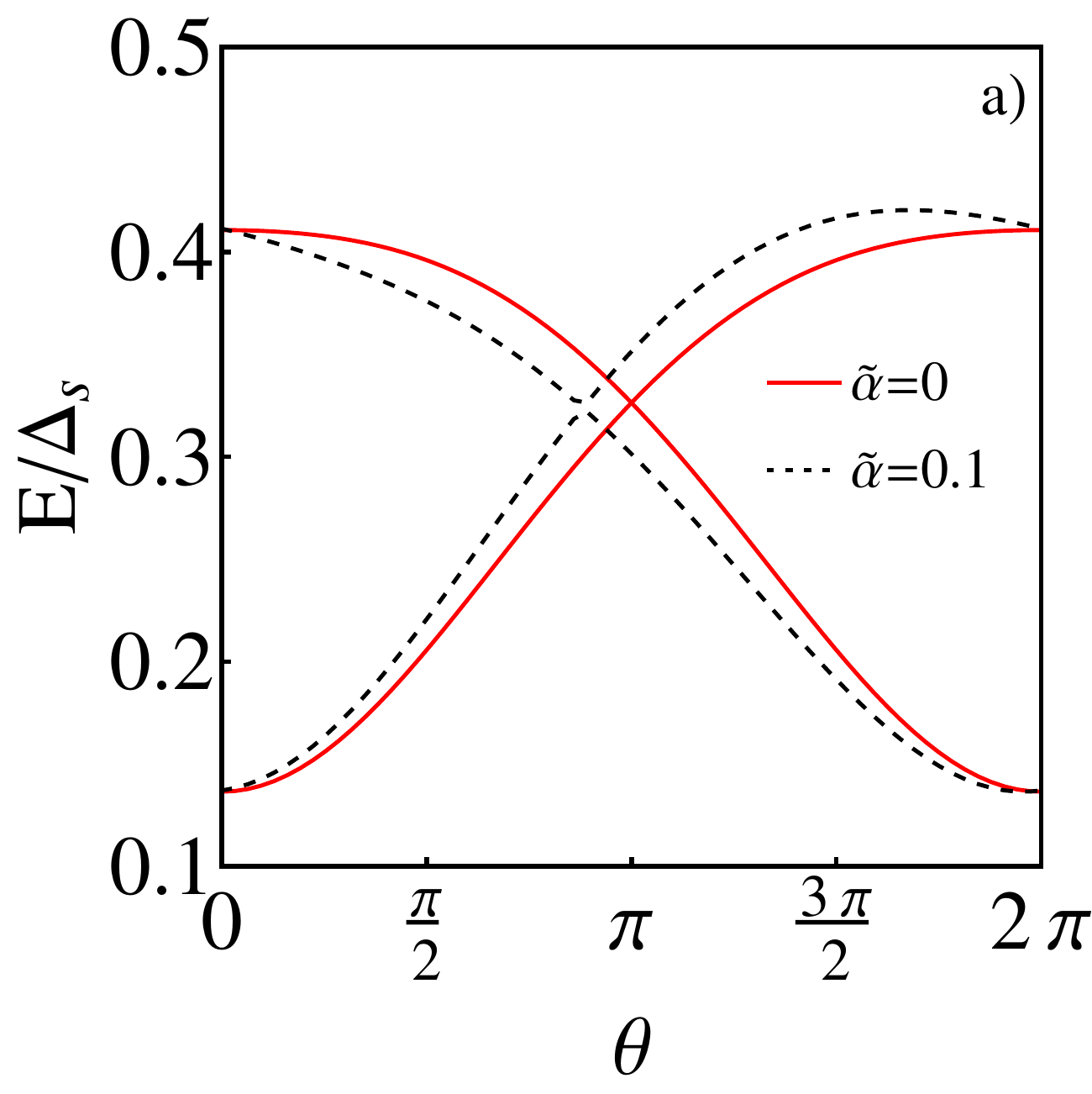}\,\,\,\,
 \includegraphics[width=3.6cm]{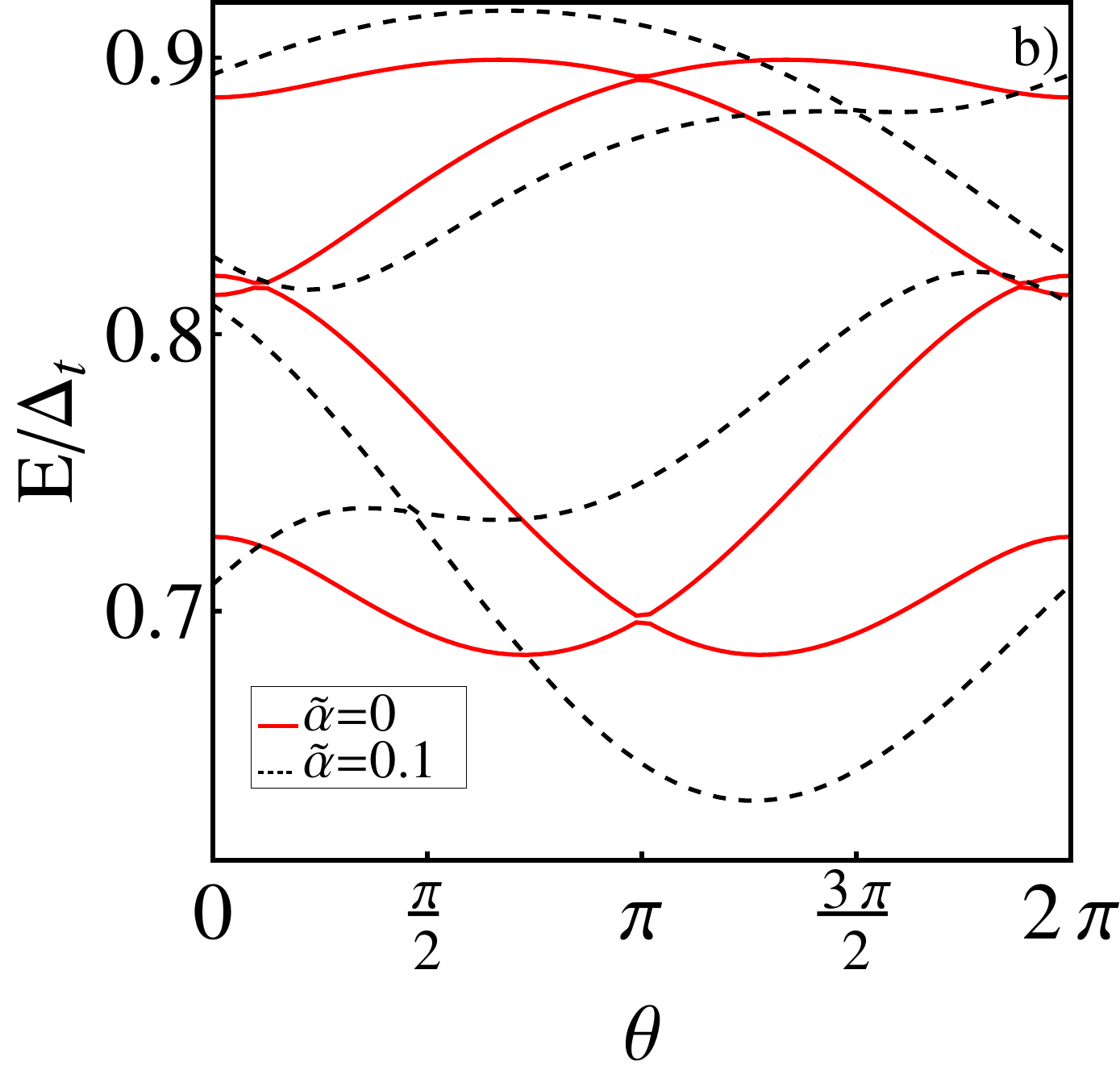}\\
 \includegraphics[width=3.8cm]{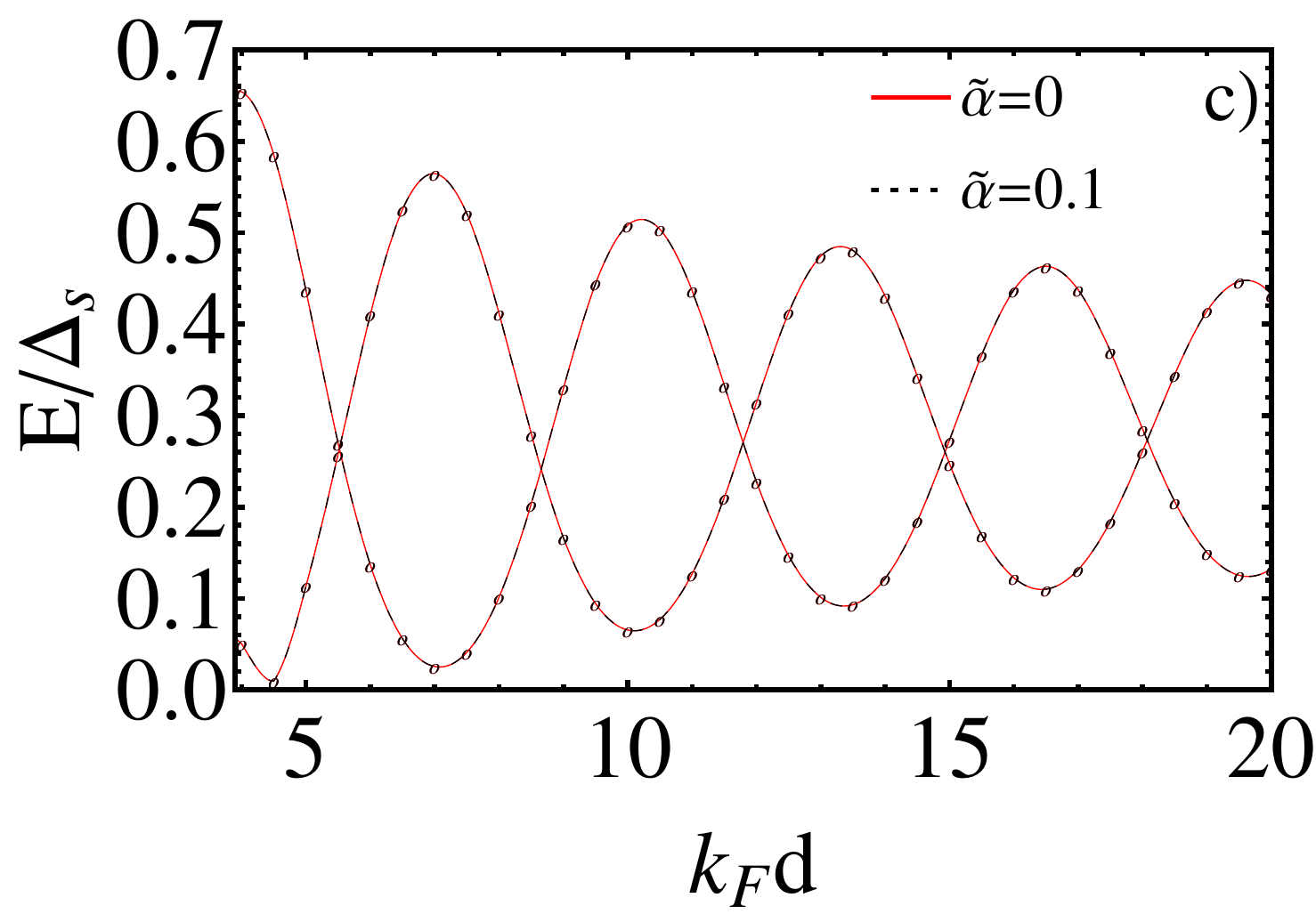}\,\,\,\,
 \includegraphics[width=3.8cm]{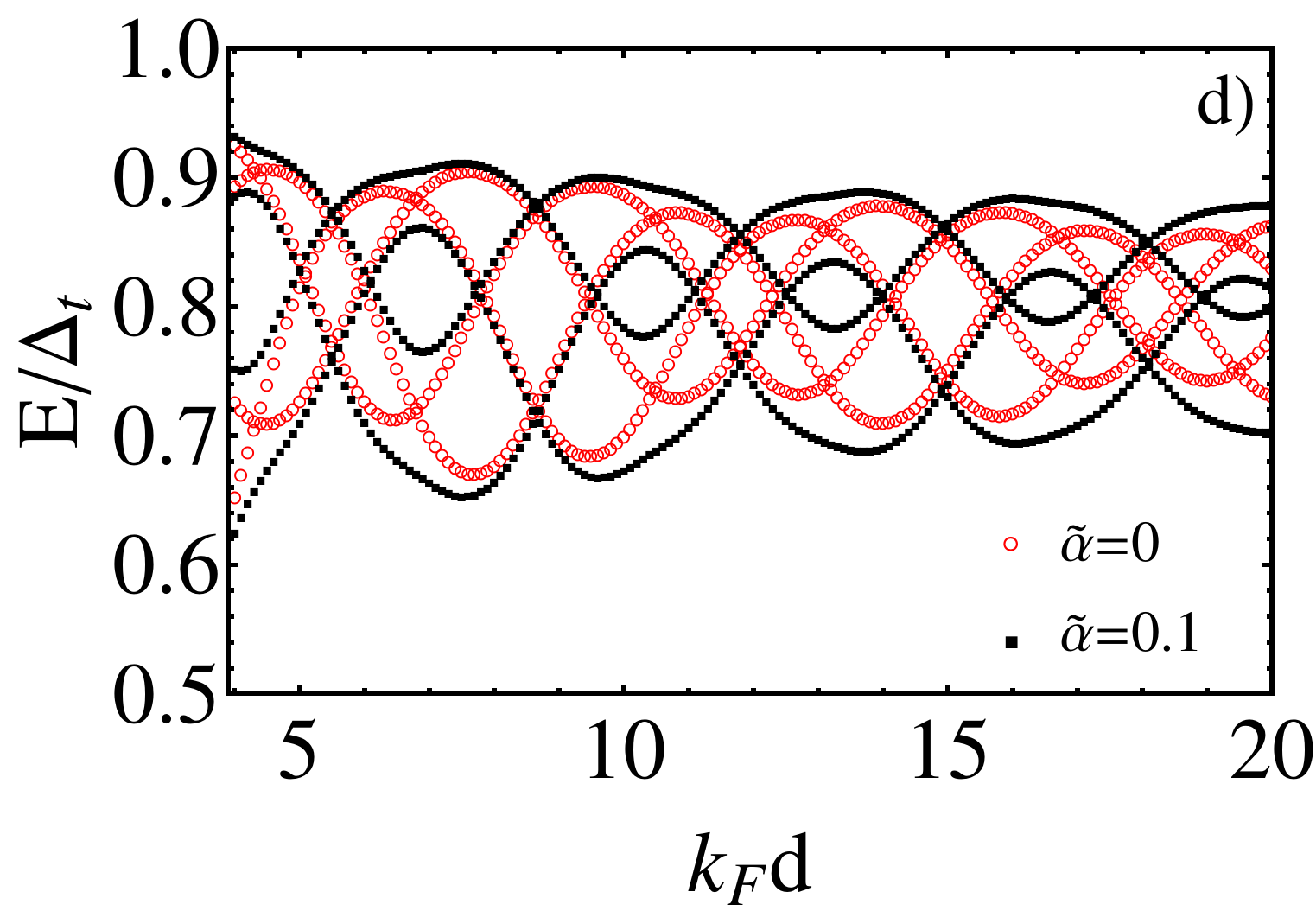}
\caption{Bound state spectrum of a magnetic impurity dimer along the $\hat{x}$ in a s-wave (a) and p-wave (b) SC. The direction of one impurity is fixed along $\hat{z}$ while the other impurity pointing in $x-z$ plane with angle $\theta$ from $\hat{z}$. Here $J_0=3/4$, $k_Fd=6$ and $\epsilon_F=1000\Delta_{s,t}$. (c) dependence of a dimer YSR spectrum on the distance $d$ between the two impurities aligned along $\hat{z}$ for an s-wave SC;
$\epsilon_F=1000\Delta_{s}$, $J_0=3/4$. (d) Same as (c) but for a p-wave SC.}
\label{fig3}
\end{figure}
The dependence of the dimer YSR spectrum on the relative angle $\theta$ between the magnetic moments of the two impurities is shown in Fig.~\ref{fig3}~(a, b). We immediately notice the following qualitative features:
 (i)   Even in the limit of no SOC, for a p-wave SC the number of energy levels is twice as large as the number of levels in an s-wave SC;
 (ii)  In the limit of no SOC for an antiferromagnetic dimer the spins of two YSR bound states are in the opposite directions such that their orthogonality leads to a level crossing at $\theta=\pi$;
 (iii) In the presence of Rashba SOC the spatial projection of a bound state spinor rotates around the $y$-axis as we move from one impurity to the other;
       as a result the crossing between levels happens at $\theta\neq\pi$;
       for an s-wave SC the two levels cross for a value of $\theta$ smaller than $\pi$ (Fig.~\ref{fig3}~(a)),
       for a p-wave SC the two lower energy states cross at $\theta<\pi$ whereas the two higher energy states
       cross at $\theta>\pi$.
					
The qualitative features listed above should be easy to test experimentally. The first feature should allow to readily identify
the symmetry, s-wave or p-wave, of the superconducting pairing in the host material, even without any tuning of the relative
angle between the magnetic moments of the two impurities. If the relative angle $\theta$ between the magnetic moments of the
two impurities is known features (ii) and (iii) allow to detect the presence of SOC and its strength. Conversely, if the
strength of the SOC is known, features (ii) and (iii) allow the determination of the relative angle $\theta$.

The properties of the system SC+dimer can be further identified by studying the dependence of the dimer YSR spectrum
on the distance $d$ between the two impurities. Figures~\ref{fig3}~(c),~(d) show the evolution of the energy levels
of the YSR spectrum with $d$, for the case of an s-wave and p-wave SC respectively.
The combination of the results presented in the panels of Fig.~\ref{fig3} makes possible to
obtain experimentally, by measuring the dependence of the dimer spectrum on the experimentally tunable parameters $\theta$ and $d$:
(i) the pairing symmetry of the SC, (ii) the strength of the impurity-host exchange couplings ($J_n)$, (iii) the strength of the SOC.


{\it Conclusions.}
We have studied the effect of spin-orbit coupling on the impurity-induced resonances in the local density of states of a 2D superconductor
for the case of a single impurity and a dimer.
Our treatment is general in that:
(i) it allows for the presence of s-wave and p-wave superconducting pairings,
(ii) it includes higher ($|l|\geq 1$) angular momentum components of the impurity potential,
(iii) it takes into account both the scalar and the magnetic part of the impurity potential.
We show that SOC mixes YSR
states with different angular momentum and therefore strongly modifies their spectrum.
In particular we find that:
(i) In the presence of SOC the parity of the particle (or hole)-like energy levels of the YSR spectrum is different in s-wave and p-wave SCs,
a fact that should allow one to identify the dominant superconducting pairing symmetry of the host material;
(ii)  By changing the direction of the magnetic moment of the impurity the fermion-parity of the lowest YSR state can be modified;
(iii) The dimer YSR spectrum oscillates as a function of the relative angle between the magnetic moments of the two impurities and their distance
      and that qualitative features of these oscillations depend on the superconducting pairing symmetry and the strength of the SOC.
These are predictions that can be tested experimentally using the scanning tunneling microscopy(STM) and
 %
%
have important implications for STM experiments trying to reveal the nature of the superconducting pairing in non-centrosymmetric superconductors. Since Pb has large SO coupling, our results shed some light on the measurements presented in Ref.~\onlinecite{ji2008}.
%

Our findings are also directly relevant to the ongoing efforts
to use magnetic atom chains placed on the surface of a superconductor with strong SOC, such as Pb,
to realize topological superconducting phases with Majorana end states~\cite{nadjperge2014}. Given that strong SOC leads to the dependence of the YSR spectrum on the direction of the atom magnetization, one might be able to control the fermion parity of the ground states (i.e. drive the topological quantum phase transition) by changing the direction of the magnetization. Furthermore, we argue that higher angular momentum impurity resonances might be important for the interpretation of the experiment~\cite{nadjperge2014} since it is not clear at the moment what is the dominant angular momentum channel determining the topological YSR band. Furthermore, we expect that the scalar potential $U_n$ and magnetic $J_n$ potential would vary at the ends of the chain, and may induce some additional in-gap states. The latter might give false positive signals in tunneling conductance measurements aimed to detect Majorana modes~\cite{nadjperge2014}.


\acknowledgements

R.L. acknowledges the hospitality of the Aspen Center for Physics supported by NSF grant \#1066293, where
part of this work was done. Y.K. is supported by Samsung Scholarship. JZ and ER acknowledge support
from ONR, Grant No. ONR-N00014-13-1-0321.

\bibliographystyle{apsrev}
\bibliography{Shiba}

\begin{thebibliography}{57}
\expandafter\ifx\csname natexlab\endcsname\relax\def\natexlab#1{#1}\fi
\expandafter\ifx\csname bibnamefont\endcsname\relax
  \def\bibnamefont#1{#1}\fi
\expandafter\ifx\csname bibfnamefont\endcsname\relax
  \def\bibfnamefont#1{#1}\fi
\expandafter\ifx\csname citenamefont\endcsname\relax
  \def\citenamefont#1{#1}\fi
\expandafter\ifx\csname url\endcsname\relax
  \def\url#1{\texttt{#1}}\fi
\expandafter\ifx\csname urlprefix\endcsname\relax\def\urlprefix{URL }\fi
\providecommand{\bibinfo}[2]{#2}
\providecommand{\eprint}[2][]{\url{#2}}

\bibitem[{\citenamefont{Nadj-Perge et~al.}(2014)\citenamefont{Nadj-Perge,
  Drozdov, Li, Chen, Jeon, Seo, MacDonald, Bernevig, and
  Yazdani}}]{nadjperge2014}
\bibinfo{author}{\bibfnamefont{S.}~\bibnamefont{Nadj-Perge}},
  \bibinfo{author}{\bibfnamefont{I.~K.} \bibnamefont{Drozdov}},
  \bibinfo{author}{\bibfnamefont{J.}~\bibnamefont{Li}},
  \bibinfo{author}{\bibfnamefont{H.}~\bibnamefont{Chen}},
  \bibinfo{author}{\bibfnamefont{S.}~\bibnamefont{Jeon}},
  \bibinfo{author}{\bibfnamefont{J.}~\bibnamefont{Seo}},
  \bibinfo{author}{\bibfnamefont{A.~H.} \bibnamefont{MacDonald}},
  \bibinfo{author}{\bibfnamefont{B.~A.} \bibnamefont{Bernevig}},
  \bibnamefont{and} \bibinfo{author}{\bibfnamefont{A.}~\bibnamefont{Yazdani}},
  \bibinfo{journal}{Science}  (\bibinfo{year}{2014}).

\bibitem[{\citenamefont{Poilblanc et~al.}(1994)\citenamefont{Poilblanc,
  Scalapino, and Hanke}}]{Poilblanc'94}
\bibinfo{author}{\bibfnamefont{D.}~\bibnamefont{Poilblanc}},
  \bibinfo{author}{\bibfnamefont{D.~J.} \bibnamefont{Scalapino}},
  \bibnamefont{and} \bibinfo{author}{\bibfnamefont{W.}~\bibnamefont{Hanke}},
  \bibinfo{journal}{Phys. Rev. Lett.} \textbf{\bibinfo{volume}{72}},
  \bibinfo{pages}{884} (\bibinfo{year}{1994}).

\bibitem[{\citenamefont{Flatt\'e and Byers}(1997{\natexlab{a}})}]{Flatte'97}
\bibinfo{author}{\bibfnamefont{M.~E.} \bibnamefont{Flatt\'e}} \bibnamefont{and}
  \bibinfo{author}{\bibfnamefont{J.~M.} \bibnamefont{Byers}},
  \bibinfo{journal}{Phys. Rev. Lett.} \textbf{\bibinfo{volume}{78}},
  \bibinfo{pages}{3761} (\bibinfo{year}{1997}{\natexlab{a}}).

\bibitem[{\citenamefont{Flatt\'e and Byers}(1997{\natexlab{b}})}]{flatte1997}
\bibinfo{author}{\bibfnamefont{M.~E.} \bibnamefont{Flatt\'e}} \bibnamefont{and}
  \bibinfo{author}{\bibfnamefont{J.~M.} \bibnamefont{Byers}},
  \bibinfo{journal}{Phys. Rev. B} \textbf{\bibinfo{volume}{56}},
  \bibinfo{pages}{11213} (\bibinfo{year}{1997}{\natexlab{b}}).

\bibitem[{\citenamefont{Salkola et~al.}(1997)\citenamefont{Salkola, Balatsky,
  and Schrieffer}}]{Salkola97}
\bibinfo{author}{\bibfnamefont{M.~I.} \bibnamefont{Salkola}},
  \bibinfo{author}{\bibfnamefont{A.~V.} \bibnamefont{Balatsky}},
  \bibnamefont{and} \bibinfo{author}{\bibfnamefont{J.~R.}
  \bibnamefont{Schrieffer}}, \bibinfo{journal}{Phys. Rev. B}
  \textbf{\bibinfo{volume}{55}}, \bibinfo{pages}{12648} (\bibinfo{year}{1997}).

\bibitem[{\citenamefont{Yazdani et~al.}(1997)\citenamefont{Yazdani, Jones,
  Lutz, Crommie, and Eigler}}]{Yazdani'97}
\bibinfo{author}{\bibfnamefont{A.}~\bibnamefont{Yazdani}},
  \bibinfo{author}{\bibfnamefont{B.~A.} \bibnamefont{Jones}},
  \bibinfo{author}{\bibfnamefont{C.~P.} \bibnamefont{Lutz}},
  \bibinfo{author}{\bibfnamefont{M.~F.} \bibnamefont{Crommie}},
  \bibnamefont{and} \bibinfo{author}{\bibfnamefont{D.~M.}
  \bibnamefont{Eigler}}, \bibinfo{journal}{Science}
  \textbf{\bibinfo{volume}{275}}, \bibinfo{pages}{1767} (\bibinfo{year}{1997}).

\bibitem[{\citenamefont{Hudson et~al.}(1999)\citenamefont{Hudson, Pan, Gupta,
  Ng, and Davis}}]{Hudson'99}
\bibinfo{author}{\bibfnamefont{E.~W.} \bibnamefont{Hudson}},
  \bibinfo{author}{\bibfnamefont{S.~H.} \bibnamefont{Pan}},
  \bibinfo{author}{\bibfnamefont{A.~K.} \bibnamefont{Gupta}},
  \bibinfo{author}{\bibfnamefont{K.-W.} \bibnamefont{Ng}}, \bibnamefont{and}
  \bibinfo{author}{\bibfnamefont{J.~C.} \bibnamefont{Davis}},
  \bibinfo{journal}{Science} \textbf{\bibinfo{volume}{285}},
  \bibinfo{pages}{88} (\bibinfo{year}{1999}).

\bibitem[{\citenamefont{Lang et~al.}(2002)\citenamefont{Lang, Madhavan,
  Hoffman, Hudson, Eisaki, Uchida, and Davis}}]{lang'02}
\bibinfo{author}{\bibfnamefont{K.}~\bibnamefont{Lang}},
  \bibinfo{author}{\bibfnamefont{V.}~\bibnamefont{Madhavan}},
  \bibinfo{author}{\bibfnamefont{J.}~\bibnamefont{Hoffman}},
  \bibinfo{author}{\bibfnamefont{E.}~\bibnamefont{Hudson}},
  \bibinfo{author}{\bibfnamefont{H.}~\bibnamefont{Eisaki}},
  \bibinfo{author}{\bibfnamefont{S.}~\bibnamefont{Uchida}}, \bibnamefont{and}
  \bibinfo{author}{\bibfnamefont{J.}~\bibnamefont{Davis}},
  \bibinfo{journal}{Nature} \textbf{\bibinfo{volume}{415}},
  \bibinfo{pages}{412} (\bibinfo{year}{2002}).

\bibitem[{\citenamefont{Morr and Stavropoulos}(2003)}]{morr2003}
\bibinfo{author}{\bibfnamefont{D.~K.} \bibnamefont{Morr}} \bibnamefont{and}
  \bibinfo{author}{\bibfnamefont{N.~A.} \bibnamefont{Stavropoulos}},
  \bibinfo{journal}{Phys. Rev. B} \textbf{\bibinfo{volume}{67}},
  \bibinfo{pages}{020502} (\bibinfo{year}{2003}).

\bibitem[{\citenamefont{Balatsky et~al.}(2006)\citenamefont{Balatsky, Vekhter,
  and Zhu}}]{balatsky2006}
\bibinfo{author}{\bibfnamefont{A.~V.} \bibnamefont{Balatsky}},
  \bibinfo{author}{\bibfnamefont{I.}~\bibnamefont{Vekhter}}, \bibnamefont{and}
  \bibinfo{author}{\bibfnamefont{J.-X.} \bibnamefont{Zhu}},
  \bibinfo{journal}{Rev. Mod. Phys.} \textbf{\bibinfo{volume}{78}},
  \bibinfo{pages}{373} (\bibinfo{year}{2006}).

\bibitem[{\citenamefont{Yu}(1965)}]{yu1965}
\bibinfo{author}{\bibfnamefont{L.}~\bibnamefont{Yu}}, \bibinfo{journal}{Acta
  Physica Sinica} \textbf{\bibinfo{volume}{21}}, \bibinfo{pages}{75}
  (\bibinfo{year}{1965}).

\bibitem[{\citenamefont{{Shiba}}(1968)}]{shiba1968}
\bibinfo{author}{\bibfnamefont{H.}~\bibnamefont{{Shiba}}},
  \bibinfo{journal}{Progress of Theoretical Physics}
  \textbf{\bibinfo{volume}{40}}, \bibinfo{pages}{435} (\bibinfo{year}{1968}).

\bibitem[{\citenamefont{{Rusinov}}(1969)}]{rusinov1969}
\bibinfo{author}{\bibfnamefont{A.~I.} \bibnamefont{{Rusinov}}},
  \bibinfo{journal}{Soviet Journal of Experimental and Theoretical Physics
  Letters} \textbf{\bibinfo{volume}{9}}, \bibinfo{pages}{85}
  (\bibinfo{year}{1969}).

\bibitem[{\citenamefont{{Nadj-Perge} et~al.}(2013)\citenamefont{{Nadj-Perge},
  {Drozdov}, {Bernevig}, and {Yazdani}}}]{nadj2013}
\bibinfo{author}{\bibfnamefont{S.}~\bibnamefont{{Nadj-Perge}}},
  \bibinfo{author}{\bibfnamefont{I.~K.} \bibnamefont{{Drozdov}}},
  \bibinfo{author}{\bibfnamefont{B.~A.} \bibnamefont{{Bernevig}}},
  \bibnamefont{and}
  \bibinfo{author}{\bibfnamefont{A.}~\bibnamefont{{Yazdani}}},
  \bibinfo{journal}{Phys. Rev. B} \textbf{\bibinfo{volume}{88}},
  \bibinfo{eid}{020407} (\bibinfo{year}{2013}).

\bibitem[{\citenamefont{{Klinovaja} et~al.}(2013)\citenamefont{{Klinovaja},
  {Stano}, {Yazdani}, and {Loss}}}]{klinovaja2013}
\bibinfo{author}{\bibfnamefont{J.}~\bibnamefont{{Klinovaja}}},
  \bibinfo{author}{\bibfnamefont{P.}~\bibnamefont{{Stano}}},
  \bibinfo{author}{\bibfnamefont{A.}~\bibnamefont{{Yazdani}}},
  \bibnamefont{and} \bibinfo{author}{\bibfnamefont{D.}~\bibnamefont{{Loss}}},
  \bibinfo{journal}{Physical Review Letters} \textbf{\bibinfo{volume}{111}},
  \bibinfo{eid}{186805} (\bibinfo{year}{2013}), \eprint{1307.1442}.

\bibitem[{\citenamefont{{Braunecker} and {Simon}}(2013)}]{braunecker2013}
\bibinfo{author}{\bibfnamefont{B.}~\bibnamefont{{Braunecker}}}
  \bibnamefont{and} \bibinfo{author}{\bibfnamefont{P.}~\bibnamefont{{Simon}}},
  \bibinfo{journal}{Physical Review Letters} \textbf{\bibinfo{volume}{111}},
  \bibinfo{eid}{147202} (\bibinfo{year}{2013}), \eprint{1307.2431}.

\bibitem[{\citenamefont{{Vazifeh} and {Franz}}(2013)}]{vazifeh2013}
\bibinfo{author}{\bibfnamefont{M.~M.} \bibnamefont{{Vazifeh}}}
  \bibnamefont{and} \bibinfo{author}{\bibfnamefont{M.}~\bibnamefont{{Franz}}},
  \bibinfo{journal}{Physical Review Letters} \textbf{\bibinfo{volume}{111}},
  \bibinfo{eid}{206802} (\bibinfo{year}{2013}), \eprint{1307.2279}.

\bibitem[{\citenamefont{{Pientka} et~al.}(2013)\citenamefont{{Pientka},
  {Glazman}, and {von Oppen}}}]{pientka2013}
\bibinfo{author}{\bibfnamefont{F.}~\bibnamefont{{Pientka}}},
  \bibinfo{author}{\bibfnamefont{L.~I.} \bibnamefont{{Glazman}}},
  \bibnamefont{and} \bibinfo{author}{\bibfnamefont{F.}~\bibnamefont{{von
  Oppen}}}, \bibinfo{journal}{\prb} \textbf{\bibinfo{volume}{88}},
  \bibinfo{eid}{155420} (\bibinfo{year}{2013}).

\bibitem[{\citenamefont{Kim et~al.}(2014)\citenamefont{Kim, Cheng, Bauer,
  Lutchyn, and Das~Sarma}}]{kim2014}
\bibinfo{author}{\bibfnamefont{Y.}~\bibnamefont{Kim}},
  \bibinfo{author}{\bibfnamefont{M.}~\bibnamefont{Cheng}},
  \bibinfo{author}{\bibfnamefont{B.}~\bibnamefont{Bauer}},
  \bibinfo{author}{\bibfnamefont{R.~M.} \bibnamefont{Lutchyn}},
  \bibnamefont{and}
  \bibinfo{author}{\bibfnamefont{S.}~\bibnamefont{Das~Sarma}},
  \bibinfo{journal}{Phys. Rev. B} \textbf{\bibinfo{volume}{90}},
  \bibinfo{pages}{060401} (\bibinfo{year}{2014}).

\bibitem[{\citenamefont{{Brydon} et~al.}(2014)\citenamefont{{Brydon}, {Hui},
  and {Sau}}}]{brydon2014}
\bibinfo{author}{\bibfnamefont{P.~M.~R.} \bibnamefont{{Brydon}}},
  \bibinfo{author}{\bibfnamefont{H.-Y.} \bibnamefont{{Hui}}}, \bibnamefont{and}
  \bibinfo{author}{\bibfnamefont{J.~D.} \bibnamefont{{Sau}}},
  \bibinfo{journal}{ArXiv e-prints}  (\bibinfo{year}{2014}),
  \eprint{1407.6345}.

\bibitem[{\citenamefont{Ebisu et~al.}(2014)\citenamefont{Ebisu, Yada, Kasai,
  and Tanaka}}]{Ebisu14}
\bibinfo{author}{\bibfnamefont{H.}~\bibnamefont{Ebisu}},
  \bibinfo{author}{\bibfnamefont{K.}~\bibnamefont{Yada}},
  \bibinfo{author}{\bibfnamefont{H.}~\bibnamefont{Kasai}}, \bibnamefont{and}
  \bibinfo{author}{\bibfnamefont{Y.}~\bibnamefont{Tanaka}},
  \bibinfo{journal}{ArXiv e-prints}  (\bibinfo{year}{2014}),
  \eprint{1410.1245}.

\bibitem[{\citenamefont{Fu and Kane}(2008)}]{Fu&Kane08}
\bibinfo{author}{\bibfnamefont{L.}~\bibnamefont{Fu}} \bibnamefont{and}
  \bibinfo{author}{\bibfnamefont{C.~L.} \bibnamefont{Kane}},
  \bibinfo{journal}{Phys.\ Rev.\ Lett.} \textbf{\bibinfo{volume}{100}},
  \bibinfo{pages}{096407} (\bibinfo{year}{2008}).

\bibitem[{\citenamefont{Fu and Kane}(2009)}]{Fu&Kane09}
\bibinfo{author}{\bibfnamefont{L.}~\bibnamefont{Fu}} \bibnamefont{and}
  \bibinfo{author}{\bibfnamefont{C.~L.} \bibnamefont{Kane}},
  \bibinfo{journal}{Phys.\ Rev.\ B} \textbf{\bibinfo{volume}{79}},
  \bibinfo{pages}{161408(R)} (\bibinfo{year}{2009}).

\bibitem[{\citenamefont{Sau et~al.}(2010)\citenamefont{Sau, Lutchyn, Tewari,
  and {Das Sarma}}}]{Sau10}
\bibinfo{author}{\bibfnamefont{J.~D.} \bibnamefont{Sau}},
  \bibinfo{author}{\bibfnamefont{R.~M.} \bibnamefont{Lutchyn}},
  \bibinfo{author}{\bibfnamefont{S.}~\bibnamefont{Tewari}}, \bibnamefont{and}
  \bibinfo{author}{\bibfnamefont{S.}~\bibnamefont{{Das Sarma}}},
  \bibinfo{journal}{Phys.\ Rev.\ Lett.} \textbf{\bibinfo{volume}{104}},
  \bibinfo{pages}{040502} (\bibinfo{year}{2010}).

\bibitem[{\citenamefont{Alicea}(2010)}]{Alicea10}
\bibinfo{author}{\bibfnamefont{J.}~\bibnamefont{Alicea}},
  \bibinfo{journal}{Phys.\ Rev.\ B} \textbf{\bibinfo{volume}{81}},
  \bibinfo{pages}{125318} (\bibinfo{year}{2010}).

\bibitem[{\citenamefont{Lutchyn et~al.}(2010)\citenamefont{Lutchyn, Sau, and
  Das~Sarma}}]{LutchynPRL10}
\bibinfo{author}{\bibfnamefont{R.~M.} \bibnamefont{Lutchyn}},
  \bibinfo{author}{\bibfnamefont{J.~D.} \bibnamefont{Sau}}, \bibnamefont{and}
  \bibinfo{author}{\bibfnamefont{S.}~\bibnamefont{Das~Sarma}},
  \bibinfo{journal}{Phys.\ Rev.\ Lett.} \textbf{\bibinfo{volume}{105}},
  \bibinfo{pages}{077001} (\bibinfo{year}{2010}).

\bibitem[{\citenamefont{Oreg et~al.}(2010)\citenamefont{Oreg, Refael, and von
  Oppen}}]{1DwiresOreg}
\bibinfo{author}{\bibfnamefont{Y.}~\bibnamefont{Oreg}},
  \bibinfo{author}{\bibfnamefont{G.}~\bibnamefont{Refael}}, \bibnamefont{and}
  \bibinfo{author}{\bibfnamefont{F.}~\bibnamefont{von Oppen}},
  \bibinfo{journal}{Phys.\ Rev.\ Lett.} \textbf{\bibinfo{volume}{105}},
  \bibinfo{pages}{177002} (\bibinfo{year}{2010}).

\bibitem[{\citenamefont{Linder et~al.}(2010)\citenamefont{Linder, Tanaka,
  Yokoyama, Sudb\o{}, and Nagaosa}}]{Linder10}
\bibinfo{author}{\bibfnamefont{J.}~\bibnamefont{Linder}},
  \bibinfo{author}{\bibfnamefont{Y.}~\bibnamefont{Tanaka}},
  \bibinfo{author}{\bibfnamefont{T.}~\bibnamefont{Yokoyama}},
  \bibinfo{author}{\bibfnamefont{A.}~\bibnamefont{Sudb\o{}}}, \bibnamefont{and}
  \bibinfo{author}{\bibfnamefont{N.}~\bibnamefont{Nagaosa}},
  \bibinfo{journal}{Phys. Rev. Lett.} \textbf{\bibinfo{volume}{104}},
  \bibinfo{pages}{067001} (\bibinfo{year}{2010}).

\bibitem[{\citenamefont{Cook and Franz}(2011)}]{MajoranaTInanowires}
\bibinfo{author}{\bibfnamefont{A.}~\bibnamefont{Cook}} \bibnamefont{and}
  \bibinfo{author}{\bibfnamefont{M.}~\bibnamefont{Franz}},
  \bibinfo{journal}{\prb} \textbf{\bibinfo{volume}{84}},
  \bibinfo{pages}{201105} (\bibinfo{year}{2011}).

\bibitem[{\citenamefont{{Lutchyn} et~al.}(2011)\citenamefont{{Lutchyn},
  {Stanescu}, and {Das Sarma}}}]{1DwiresLutchyn2}
\bibinfo{author}{\bibfnamefont{R.~M.} \bibnamefont{{Lutchyn}}},
  \bibinfo{author}{\bibfnamefont{T.~D.} \bibnamefont{{Stanescu}}},
  \bibnamefont{and} \bibinfo{author}{\bibfnamefont{S.}~\bibnamefont{{Das
  Sarma}}}, \bibinfo{journal}{\prl} \textbf{\bibinfo{volume}{106}},
  \bibinfo{eid}{127001} (\bibinfo{year}{2011}).

\bibitem[{\citenamefont{Duckheim and Brouwer}(2011)}]{Duckheim'11}
\bibinfo{author}{\bibfnamefont{M.}~\bibnamefont{Duckheim}} \bibnamefont{and}
  \bibinfo{author}{\bibfnamefont{P.~W.} \bibnamefont{Brouwer}},
  \bibinfo{journal}{Phys. Rev. B} \textbf{\bibinfo{volume}{83}},
  \bibinfo{pages}{054513} (\bibinfo{year}{2011}).

\bibitem[{\citenamefont{Choy et~al.}(2011)\citenamefont{Choy, Edge, Akhmerov,
  and Beenakker}}]{Choy'11}
\bibinfo{author}{\bibfnamefont{T.-P.} \bibnamefont{Choy}},
  \bibinfo{author}{\bibfnamefont{J.~M.} \bibnamefont{Edge}},
  \bibinfo{author}{\bibfnamefont{A.~R.} \bibnamefont{Akhmerov}},
  \bibnamefont{and} \bibinfo{author}{\bibfnamefont{C.~W.~J.}
  \bibnamefont{Beenakker}}, \bibinfo{journal}{Phys. Rev. B}
  \textbf{\bibinfo{volume}{84}}, \bibinfo{pages}{195442}
  (\bibinfo{year}{2011}).

\bibitem[{\citenamefont{Chung et~al.}(2011)\citenamefont{Chung, Zhang, Qi, and
  Zhang}}]{SB'11}
\bibinfo{author}{\bibfnamefont{S.~B.} \bibnamefont{Chung}},
  \bibinfo{author}{\bibfnamefont{H.-J.} \bibnamefont{Zhang}},
  \bibinfo{author}{\bibfnamefont{X.-L.} \bibnamefont{Qi}}, \bibnamefont{and}
  \bibinfo{author}{\bibfnamefont{S.-C.} \bibnamefont{Zhang}},
  \bibinfo{journal}{Phys. Rev. B} \textbf{\bibinfo{volume}{84}},
  \bibinfo{pages}{060510} (\bibinfo{year}{2011}).

\bibitem[{\citenamefont{Kjaergaard et~al.}(2012)\citenamefont{Kjaergaard,
  W\"olms, and Flensberg}}]{Flensberg'12}
\bibinfo{author}{\bibfnamefont{M.}~\bibnamefont{Kjaergaard}},
  \bibinfo{author}{\bibfnamefont{K.}~\bibnamefont{W\"olms}}, \bibnamefont{and}
  \bibinfo{author}{\bibfnamefont{K.}~\bibnamefont{Flensberg}},
  \bibinfo{journal}{Phys. Rev. B} \textbf{\bibinfo{volume}{85}},
  \bibinfo{pages}{020503} (\bibinfo{year}{2012}).

\bibitem[{\citenamefont{Potter and Lee}(2012)}]{Potter'12}
\bibinfo{author}{\bibfnamefont{A.~C.} \bibnamefont{Potter}} \bibnamefont{and}
  \bibinfo{author}{\bibfnamefont{P.~A.} \bibnamefont{Lee}},
  \bibinfo{journal}{Phys. Rev. B} \textbf{\bibinfo{volume}{85}},
  \bibinfo{pages}{094516} (\bibinfo{year}{2012}).

\bibitem[{\citenamefont{Martin and Morpurgo}(2012)}]{Martin'12}
\bibinfo{author}{\bibfnamefont{I.}~\bibnamefont{Martin}} \bibnamefont{and}
  \bibinfo{author}{\bibfnamefont{A.~F.} \bibnamefont{Morpurgo}},
  \bibinfo{journal}{Phys. Rev. B} \textbf{\bibinfo{volume}{85}},
  \bibinfo{pages}{144505} (\bibinfo{year}{2012}).

\bibitem[{\citenamefont{{Sau} and {Sarma}}(2012)}]{SauNature'12}
\bibinfo{author}{\bibfnamefont{J.~D.} \bibnamefont{{Sau}}} \bibnamefont{and}
  \bibinfo{author}{\bibfnamefont{S.~D.} \bibnamefont{{Sarma}}},
  \bibinfo{journal}{Nat. Commun.} \textbf{\bibinfo{volume}{3}}
  (\bibinfo{year}{2012}).

\bibitem[{\citenamefont{{Mourik} et~al.}(2012)\citenamefont{{Mourik}, {Zuo},
  {Frolov}, {Plissard}, {Bakkers}, and {Kouwenhoven}}}]{Mourik2012}
\bibinfo{author}{\bibfnamefont{V.}~\bibnamefont{{Mourik}}},
  \bibinfo{author}{\bibfnamefont{K.}~\bibnamefont{{Zuo}}},
  \bibinfo{author}{\bibfnamefont{S.~M.} \bibnamefont{{Frolov}}},
  \bibinfo{author}{\bibfnamefont{S.~R.} \bibnamefont{{Plissard}}},
  \bibinfo{author}{\bibfnamefont{E.~P.~A.~M.} \bibnamefont{{Bakkers}}},
  \bibnamefont{and} \bibinfo{author}{\bibfnamefont{L.~P.}
  \bibnamefont{{Kouwenhoven}}}, \bibinfo{journal}{Science}
  \textbf{\bibinfo{volume}{336}}, \bibinfo{pages}{1003} (\bibinfo{year}{2012}).

\bibitem[{\citenamefont{Rokhinson et~al.}(2012)\citenamefont{Rokhinson, Liu,
  and Furdyna}}]{Rokhinson2012}
\bibinfo{author}{\bibfnamefont{L.~P.} \bibnamefont{Rokhinson}},
  \bibinfo{author}{\bibfnamefont{X.~Y.} \bibnamefont{Liu}}, \bibnamefont{and}
  \bibinfo{author}{\bibfnamefont{J.~K.} \bibnamefont{Furdyna}},
  \bibinfo{journal}{Nature Phys.} \textbf{\bibinfo{volume}{8}},
  \bibinfo{pages}{795} (\bibinfo{year}{2012}).

\bibitem[{\citenamefont{Das et~al.}(2012)\citenamefont{Das, Ronen, Most, Oreg,
  Heiblum, and Shtrikman}}]{Das2012}
\bibinfo{author}{\bibfnamefont{A.}~\bibnamefont{Das}},
  \bibinfo{author}{\bibfnamefont{Y.}~\bibnamefont{Ronen}},
  \bibinfo{author}{\bibfnamefont{Y.}~\bibnamefont{Most}},
  \bibinfo{author}{\bibfnamefont{Y.}~\bibnamefont{Oreg}},
  \bibinfo{author}{\bibfnamefont{M.}~\bibnamefont{Heiblum}}, \bibnamefont{and}
  \bibinfo{author}{\bibfnamefont{H.}~\bibnamefont{Shtrikman}},
  \bibinfo{journal}{Nature Phys.} \textbf{\bibinfo{volume}{8}},
  \bibinfo{pages}{887} (\bibinfo{year}{2012}).

\bibitem[{\citenamefont{{Deng} et~al.}(2012)\citenamefont{{Deng}, {Yu},
  {Huang}, {Larsson}, {Caroff}, and {Xu}}}]{Deng2012}
\bibinfo{author}{\bibfnamefont{M.~T.} \bibnamefont{{Deng}}},
  \bibinfo{author}{\bibfnamefont{C.~L.} \bibnamefont{{Yu}}},
  \bibinfo{author}{\bibfnamefont{G.~Y.} \bibnamefont{{Huang}}},
  \bibinfo{author}{\bibfnamefont{M.}~\bibnamefont{{Larsson}}},
  \bibinfo{author}{\bibfnamefont{P.}~\bibnamefont{{Caroff}}}, \bibnamefont{and}
  \bibinfo{author}{\bibfnamefont{H.~Q.} \bibnamefont{{Xu}}},
  \bibinfo{journal}{Nano Lett.} \textbf{\bibinfo{volume}{12}},
  \bibinfo{pages}{6414} (\bibinfo{year}{2012}).

\bibitem[{\citenamefont{{Finck} et~al.}(2013)\citenamefont{{Finck}, {Van
  Harlingen}, {Mohseni}, {Jung}, and {Li}}}]{Fink2012}
\bibinfo{author}{\bibfnamefont{A.~D.~K.} \bibnamefont{{Finck}}},
  \bibinfo{author}{\bibfnamefont{D.~J.} \bibnamefont{{Van Harlingen}}},
  \bibinfo{author}{\bibfnamefont{P.~K.} \bibnamefont{{Mohseni}}},
  \bibinfo{author}{\bibfnamefont{K.}~\bibnamefont{{Jung}}}, \bibnamefont{and}
  \bibinfo{author}{\bibfnamefont{X.}~\bibnamefont{{Li}}},
  \bibinfo{journal}{\prl} \textbf{\bibinfo{volume}{110}},
  \bibinfo{pages}{126406} (\bibinfo{year}{2013}).

\bibitem[{\citenamefont{Churchill et~al.}(2013)\citenamefont{Churchill, Fatemi,
  Grove-Rasmussen, Deng, Caroff, Xu, and Marcus}}]{Churchill2013}
\bibinfo{author}{\bibfnamefont{H.~O.~H.} \bibnamefont{Churchill}},
  \bibinfo{author}{\bibfnamefont{V.}~\bibnamefont{Fatemi}},
  \bibinfo{author}{\bibfnamefont{K.}~\bibnamefont{Grove-Rasmussen}},
  \bibinfo{author}{\bibfnamefont{M.~T.} \bibnamefont{Deng}},
  \bibinfo{author}{\bibfnamefont{P.}~\bibnamefont{Caroff}},
  \bibinfo{author}{\bibfnamefont{H.~Q.} \bibnamefont{Xu}}, \bibnamefont{and}
  \bibinfo{author}{\bibfnamefont{C.~M.} \bibnamefont{Marcus}},
  \bibinfo{journal}{\prb} \textbf{\bibinfo{volume}{87}},
  \bibinfo{pages}{241401} (\bibinfo{year}{2013}).

\bibitem[{\citenamefont{{Hui} et~al.}(2014)\citenamefont{{Hui}, {Brydon},
  {Sau}, {Tewari}, and {Das Sarma}}}]{hui2014}
\bibinfo{author}{\bibfnamefont{H.-Y.} \bibnamefont{{Hui}}},
  \bibinfo{author}{\bibfnamefont{P.~M.~R.} \bibnamefont{{Brydon}}},
  \bibinfo{author}{\bibfnamefont{J.~D.} \bibnamefont{{Sau}}},
  \bibinfo{author}{\bibfnamefont{S.}~\bibnamefont{{Tewari}}}, \bibnamefont{and}
  \bibinfo{author}{\bibfnamefont{S.}~\bibnamefont{{Das Sarma}}},
  \bibinfo{journal}{ArXiv e-prints}  (\bibinfo{year}{2014}),
  \eprint{1407.7519}.

\bibitem[{\citenamefont{Li et~al.}(2014)\citenamefont{Li, Chen, Drozdov,
  Yazdani, Bernevig, and MacDonald}}]{Li14}
\bibinfo{author}{\bibfnamefont{J.}~\bibnamefont{Li}},
  \bibinfo{author}{\bibfnamefont{H.}~\bibnamefont{Chen}},
  \bibinfo{author}{\bibfnamefont{I.~K.} \bibnamefont{Drozdov}},
  \bibinfo{author}{\bibfnamefont{A.}~\bibnamefont{Yazdani}},
  \bibinfo{author}{\bibfnamefont{B.~A.} \bibnamefont{Bernevig}},
  \bibnamefont{and} \bibinfo{author}{\bibfnamefont{A.~H.}
  \bibnamefont{MacDonald}}, \bibinfo{journal}{Phys. Rev. B}
  \textbf{\bibinfo{volume}{90}}, \bibinfo{pages}{235433}
  (\bibinfo{year}{2014}).

\bibitem[{\citenamefont{Gor'kov and Rashba}(2001)}]{Gorkov01}
\bibinfo{author}{\bibfnamefont{L.~P.} \bibnamefont{Gor'kov}} \bibnamefont{and}
  \bibinfo{author}{\bibfnamefont{E.~I.} \bibnamefont{Rashba}},
  \bibinfo{journal}{Phys. Rev. Lett.} \textbf{\bibinfo{volume}{87}},
  \bibinfo{pages}{037004} (\bibinfo{year}{2001}).

\bibitem[{\citenamefont{Kunz and Ginsberg}(1980)}]{Kunz'80}
\bibinfo{author}{\bibfnamefont{A.~B.} \bibnamefont{Kunz}} \bibnamefont{and}
  \bibinfo{author}{\bibfnamefont{D.~M.} \bibnamefont{Ginsberg}},
  \bibinfo{journal}{Phys. Rev. B} \textbf{\bibinfo{volume}{22}},
  \bibinfo{pages}{3165} (\bibinfo{year}{1980}).

\bibitem[{\citenamefont{Ji et~al.}(2008)\citenamefont{Ji, Zhang, Fu, Chen, Ma,
  Li, Duan, Jia, and Xue}}]{ji2008}
\bibinfo{author}{\bibfnamefont{S.-H.} \bibnamefont{Ji}},
  \bibinfo{author}{\bibfnamefont{T.}~\bibnamefont{Zhang}},
  \bibinfo{author}{\bibfnamefont{Y.-S.} \bibnamefont{Fu}},
  \bibinfo{author}{\bibfnamefont{X.}~\bibnamefont{Chen}},
  \bibinfo{author}{\bibfnamefont{X.-C.} \bibnamefont{Ma}},
  \bibinfo{author}{\bibfnamefont{J.}~\bibnamefont{Li}},
  \bibinfo{author}{\bibfnamefont{W.-H.} \bibnamefont{Duan}},
  \bibinfo{author}{\bibfnamefont{J.-F.} \bibnamefont{Jia}}, \bibnamefont{and}
  \bibinfo{author}{\bibfnamefont{Q.-K.} \bibnamefont{Xue}},
  \bibinfo{journal}{Phys. Rev. Lett.} \textbf{\bibinfo{volume}{100}},
  \bibinfo{pages}{226801} (\bibinfo{year}{2008}).

\bibitem[{\citenamefont{Grothe et~al.}(2012)\citenamefont{Grothe, Chi, Dosanjh,
  Liang, Hardy, Burke, Bonn, and Pennec}}]{Grothe'12}
\bibinfo{author}{\bibfnamefont{S.}~\bibnamefont{Grothe}},
  \bibinfo{author}{\bibfnamefont{S.}~\bibnamefont{Chi}},
  \bibinfo{author}{\bibfnamefont{P.}~\bibnamefont{Dosanjh}},
  \bibinfo{author}{\bibfnamefont{R.}~\bibnamefont{Liang}},
  \bibinfo{author}{\bibfnamefont{W.~N.} \bibnamefont{Hardy}},
  \bibinfo{author}{\bibfnamefont{S.~A.} \bibnamefont{Burke}},
  \bibinfo{author}{\bibfnamefont{D.~A.} \bibnamefont{Bonn}}, \bibnamefont{and}
  \bibinfo{author}{\bibfnamefont{Y.}~\bibnamefont{Pennec}},
  \bibinfo{journal}{Phys. Rev. B} \textbf{\bibinfo{volume}{86}},
  \bibinfo{pages}{174503} (\bibinfo{year}{2012}).

\bibitem[{\citenamefont{Frigeri et~al.}(2004)\citenamefont{Frigeri, Agterberg,
  Koga, and Sigrist}}]{Frigeri04}
\bibinfo{author}{\bibfnamefont{P.~A.} \bibnamefont{Frigeri}},
  \bibinfo{author}{\bibfnamefont{D.~F.} \bibnamefont{Agterberg}},
  \bibinfo{author}{\bibfnamefont{A.}~\bibnamefont{Koga}}, \bibnamefont{and}
  \bibinfo{author}{\bibfnamefont{M.}~\bibnamefont{Sigrist}},
  \bibinfo{journal}{Phys. Rev. Lett.} \textbf{\bibinfo{volume}{92}},
  \bibinfo{pages}{097001} (\bibinfo{year}{2004}).

\bibitem[{sm()}]{sm}
\bibinfo{note}{See supplementary material.}

\bibitem[{\citenamefont{Sato and Fujimoto}(2009)}]{Sato09}
\bibinfo{author}{\bibfnamefont{M.}~\bibnamefont{Sato}} \bibnamefont{and}
  \bibinfo{author}{\bibfnamefont{S.}~\bibnamefont{Fujimoto}},
  \bibinfo{journal}{Phys. Rev. B} \textbf{\bibinfo{volume}{79}},
  \bibinfo{pages}{094504} (\bibinfo{year}{2009}).

\bibitem[{\citenamefont{Tewari et~al.}(2011)\citenamefont{Tewari, Stanescu,
  Sau, and Sarma}}]{Tewari'11}
\bibinfo{author}{\bibfnamefont{S.}~\bibnamefont{Tewari}},
  \bibinfo{author}{\bibfnamefont{T.~D.} \bibnamefont{Stanescu}},
  \bibinfo{author}{\bibfnamefont{J.~D.} \bibnamefont{Sau}}, \bibnamefont{and}
  \bibinfo{author}{\bibfnamefont{S.~D.} \bibnamefont{Sarma}},
  \bibinfo{journal}{New Journal of Physics} \textbf{\bibinfo{volume}{13}},
  \bibinfo{pages}{065004} (\bibinfo{year}{2011}).

\bibitem[{\citenamefont{Sau and Demler}(2013)}]{SauDemler'13}
\bibinfo{author}{\bibfnamefont{J.~D.} \bibnamefont{Sau}} \bibnamefont{and}
  \bibinfo{author}{\bibfnamefont{E.}~\bibnamefont{Demler}},
  \bibinfo{journal}{Phys. Rev. B} \textbf{\bibinfo{volume}{88}},
  \bibinfo{pages}{205402} (\bibinfo{year}{2013}).

\bibitem[{\citenamefont{Wang and Wang}(2004)}]{Wang04}
\bibinfo{author}{\bibfnamefont{Q.-H.} \bibnamefont{Wang}} \bibnamefont{and}
  \bibinfo{author}{\bibfnamefont{Z.~D.} \bibnamefont{Wang}},
  \bibinfo{journal}{Phys. Rev. B} \textbf{\bibinfo{volume}{69}},
  \bibinfo{pages}{092502} (\bibinfo{year}{2004}).

\bibitem[{\citenamefont{Liu and Eremin}(2008)}]{Liu08}
\bibinfo{author}{\bibfnamefont{B.}~\bibnamefont{Liu}} \bibnamefont{and}
  \bibinfo{author}{\bibfnamefont{I.}~\bibnamefont{Eremin}},
  \bibinfo{journal}{Phys. Rev. B} \textbf{\bibinfo{volume}{78}},
  \bibinfo{pages}{014518} (\bibinfo{year}{2008}).

\bibitem[{\citenamefont{Nagai et~al.}(2014)\citenamefont{Nagai, Ota, and
  Machida}}]{Nagai14}
\bibinfo{author}{\bibfnamefont{Y.}~\bibnamefont{Nagai}},
  \bibinfo{author}{\bibfnamefont{Y.}~\bibnamefont{Ota}}, \bibnamefont{and}
  \bibinfo{author}{\bibfnamefont{M.}~\bibnamefont{Machida}},
  \bibinfo{journal}{Phys. Rev. B} \textbf{\bibinfo{volume}{89}},
  \bibinfo{pages}{214506} (\bibinfo{year}{2014}).

\end{thebibliography}

\newpage

\onecolumngrid
\begin{center}

{\bf\Large{Supplemental Material for ``Impurity-induced bound states in superconductors with spin-orbit coupling''}}

\end{center}
\vspace{1cm}
\setcounter{equation}{0}
\renewcommand{\theequation}{S\arabic{equation}}
\setcounter{figure}{0}
\renewcommand{\thefigure}{S\arabic{figure}}

\section{Momentum averaged Green's function}

The momentum integral of Green's functions $\overline{\hat{G}^{ij}(E,\theta)}$ can be derived by splitting $\hat{G}(E,\vec{p})$ into two branches $\hat{G}^{\pm}(E,\vec{p})$ and changing the integral over the momentum to an integral over energy dispersion $\xi_\pm$ for each branch:
\ba
\overline{\hat G^{ij}(E,\theta)}&=&\frac{1}{2}(\overline{\hat G^{+,ij}(E,\theta)}+\overline{\hat G^{-,ij}(E,\theta)})\\
\frac{1}{2}\overline{\hat G^{\lambda,ij}(E,\theta)}&=& \frac{1}{2}\int_0^\infty  \frac{dp}{2\pi} \,p e^{-ix_{ij}p\cos\theta}\, \hat{G^\lambda}(E,\vec{p}) \\
&\approx& \frac{\nu_\lambda}{2} \int_{-\Lambda}^\Lambda d\xi_\lambda \, e^{-ix_{ij}p_\lambda(\xi_\lambda)\cos\theta}\hat{G^\lambda}(E,\xi_\lambda,\theta)
\ea
where $x_{ij}=x_i-x_j$, $\lambda=\pm$, $\nu_\lambda=\frac{m}{2\pi}(1-\frac{\lambda\tilde{\alpha}}{\sqrt{1+\tilde{\alpha}^2}})$, $\tilde{\alpha}=m\alpha/p_F$, $p_F=\sqrt{2m \varepsilon_F}$, $p_\lambda(\xi)=p_{F\lambda}+\xi /v_F$, $\,p_{F\lambda}=p_F(\sqrt{1+\tilde{\alpha}^2}-\lambda\tilde{\alpha})$, $v_F= p_F\sqrt{1+\tilde{\alpha}^2}/m$ and $\Lambda$ is a cutoff. Assuming the most of the contributions for $\overline{\hat{G}^{ij}(E,\theta)}$ comes from near the Fermi surface $p\sim p_F$, we substitute $\Delta_\pm$ with $\tilde{\Delta}_\pm=\Delta_s\pm\Delta_t$. The analytic forms of $\overline{\hat{G}^{\lambda,ij}_n(E,\theta)}$ can be derived in the limit $\Lambda\rightarrow\infty$ by using the following integrals:
\ba
I_A^\lambda(x,\theta)&=&\frac{\nu_\lambda}{2}\int_{-\infty}^\infty d\xi_\lambda \frac{e^{ip_\lambda(\xi_\lambda)x \cos \theta }}{E^2-\xi_\lambda^2-\Delta^2}\\
&=&-\frac{\pi\nu_\lambda}{2\sqrt{\Delta^2-E^2}}\exp[ip_{F\lambda}x\cos\theta-\frac{\sqrt{\Delta^2-E^2}}{v_F}\,|x \cos \theta |]
\ea
\ba
I_B^\lambda(x,\theta)&=&\frac{\nu_\lambda}{2}\int_{-\infty}^\infty d\xi_\lambda \frac{\xi_\lambda e^{ip_\lambda(\xi_\lambda) x \cos\theta }}{E^2-\xi_\lambda^2-\Delta^2}\\
&=&
\begin{cases}
-\frac{i\pi\nu_\lambda\sgn(x\cos\theta)}{2}\exp[ip_{F\lambda}x\cos\theta-\frac{\sqrt{\Delta^2-E^2}}{v_F}\,|x \cos \theta |], & x\neq0\\
0,  & x=0
\end{cases}
\ea
Their angular momentum components are defined by
\ba
I_{A,n}^\lambda(x)&=&\frac{1}{2\pi}\int_0^{2\pi} I_A^\lambda(x,\theta)e^{-in\theta}d\theta,\\
I_{B,n}^\lambda(x\neq 0)&=&\frac{1}{2\pi}\int_0^{2\pi} I_B^\lambda(x\neq 0,\theta)e^{-in\theta}d\theta.
\ea
By changing the integration variable $\theta\rightarrow-\theta$ we
obtain  the identity $I^\lambda_{A/B,n}=I^\lambda_{A/B,-n}$.
The results of the above integrals for $|n|\leq 3$ can be written as
\ba
I_{A,0}^\lambda(x)
&=&-\frac{\pi\nu_\lambda}{2\sqrt{\Delta^2-E^2}}\re\left[J_0(\beta_\lambda(|x|))+iH_0(\beta_\lambda(|x|))\right],\\
I_{A,\pm 1}^\lambda(x)
&=&-\frac{i\pi\nu_\lambda\sgn(x)}{2\sqrt{\Delta^2-E^2}}\im\left[iJ_1(\beta_\lambda(|x|))+H_{-1}(\beta_\lambda(|x|))\right], \\
I_{A, \pm2}^\lambda(x)&=&-\frac{\pi\nu_\lambda}{2\sqrt{\Delta^2-E^2}}\re\left[-J_2(\beta_\lambda(|x|))+iH_0(\beta_\lambda(|x|))-\frac{2iH_1(\beta_\lambda(|x|))}{\beta_\lambda(|x|)} \right],\\
I_{A, \pm3}^\lambda(x)&=&-\frac{i\pi\nu_\lambda\sgn(x)}{2\sqrt{\Delta^2-E^2}}\im\left[-\frac{8}{3\pi}-iJ_3(\beta_\lambda(|x|))+H_{-1}(\beta_\lambda(|x|))+\frac{4iH_2(\beta_\lambda(|x|))}{\beta_\lambda(|x|)} \right]
\ea
and for non-zero $x$
\ba
I_{B,0}^\lambda(x)
&=&\frac{\pi\nu_\lambda}{2}\im\left[J_0(\beta_\lambda(|x|))+iH_0(\beta_\lambda(|x|))\right],\\
I_{B,\pm 1}^\lambda(x)
&=&-\frac{i\pi\nu_\lambda\sgn(x)}{2}\re\left[iJ_1(\beta_\lambda(|x|))+H_{-1}(\beta_\lambda(|x|))\right]\\
I_{B, \pm2}^\lambda(x)&=&\frac{\pi\nu_\lambda}{2}\im\left[-J_2(\beta_\lambda(|x|))+iH_0(\beta_\lambda(|x|))-\frac{2iH_1(\beta_\lambda(|x|))}{\beta_\lambda(|x|)} \right],\\
I_{B, \pm3}^\lambda(x)&=&-\frac{i\pi\nu_\lambda\sgn(x)}{2}\re\left[-\frac{8}{3\pi}-iJ_3(\beta_\lambda(|x|))+H_{-1}(\beta_\lambda(|x|))+\frac{4iH_2(\beta_\lambda(|x|))}{\beta_\lambda(|x|)} \right]
\ea
where $\beta_\lambda(x)=p_{F\lambda}x+i\frac{\sqrt{\Delta^2-E^2}}{v_F}|x|\approx p_{F\lambda}x$ for $\Delta/\epsilon_F\ll1$, $J_n(z)$ and $H_n(z)$ are Bessel and Struve functions of order $n$. Note that $I^\lambda_{B,n}(0)=0$.
Using the above results for $x=0$, we get
\ba
\overline{\hat{G}^{ii}_0(E)}&=&-\frac{\pi\nu_+}{2\sqrt{\Delta_+^2-E^2}}\left( \begin{array}{cccc}
E & 0 & \Delta_+  &0 \\
0 & E & 0 & \Delta_+ \\
\Delta_+ & 0 &E &0 \\
0& \Delta_+ &0  &E
\end{array}\right)-\frac{\pi\nu_-}{2\sqrt{\Delta_-^2-E^2}}\left( \begin{array}{cccc}
E & 0 & \Delta_-  &0 \\
0 & E & 0 & \Delta_- \\
\Delta_- & 0 &E &0 \\
0& \Delta_- &0  &E
\end{array}\right),\\
\overline{\hat{G}^{ii}_1(E)}&=&-\frac{\pi\nu_+}{2\sqrt{\Delta_+^2-E^2}}\left( \begin{array}{cccc}
0 & 0 & 0 & 0 \\
-iE & 0 & -i\Delta_+ & 0 \\
0 & 0 & 0 & 0 \\
-i\Delta_+ & 0 & -iE & 0
\end{array}\right)+\frac{\pi\nu_-}{2\sqrt{\Delta_-^2-E^2}}\left( \begin{array}{cccc}
0 & 0 & 0 & 0 \\
-iE & 0 & -i\Delta_- & 0 \\
0 & 0 & 0 & 0 \\
-i\Delta_- & 0 & -iE & 0
\end{array}\right),\\
\overline{\hat{G}^{ii}_{-1}(E)}&=&-\frac{\pi\nu_+}{2\sqrt{\Delta_+^2-E^2}}\left( \begin{array}{cccc}
0 & iE & 0 & i\Delta_+ \\
0 & 0 & 0 & 0 \\
0 & i\Delta_+ & 0 & iE \\
0 & 0 & 0 & 0
\end{array}\right)+\frac{\pi\nu_-}{2\sqrt{\Delta_-^2-E^2}}\left( \begin{array}{cccc}
0 & iE & 0 & i\Delta_- \\
0 & 0 & 0 & 0 \\
0 & i\Delta_- & 0 & iE \\
0 & 0 & 0 & 0
\end{array}\right),\\
\overline{\hat G^{ii}_{|n|>1}(E)}&=&0.
\ea

For $x\neq 0$,
\be
\overline{\hat{G}^{ij}_n(E)}=\frac{1}{2}\left(\overline{\hat{G}^{+,ij}_n(E)}+\overline{\hat{G}^{-,ij}_n(E)}\right)
\ee
\be
\frac{1}{2}\overline{\hat{G}^{\lambda,ij}_n(E)}=
\left(\begin{array}{cccc}
EI^\lambda_{A,n}+I^\lambda_{B,n} & \lambda i(EI^\lambda_{A,n+1}+I^\lambda_{B,n+1}) & \Delta I^\lambda_{A,n} & \lambda i \Delta I^\lambda_{A,n+1}\\
 -\lambda i(EI^\lambda_{A,n-1}+I^\lambda_{B,n-1}) & EI^\lambda_{A,n}+I^\lambda_{B,n} & -\lambda i \Delta I^\lambda_{A,n-1} & \Delta I^\lambda_{A,n}\\
\Delta I^\lambda_{A,n} & \lambda i \Delta I^\lambda_{A,n+1} & EI^\lambda_{A,n}-I^\lambda_{B,n} & i\lambda(EI^\lambda_{A,n+1}-I^\lambda_{B,n+1})\\
 -\lambda i \Delta I^\lambda_{A,n-1} & \Delta I^\lambda_{A,n} & -\lambda i(EI^\lambda_{A,n-1}-I^\lambda_{B,n-1}) & EI^\lambda_{A,n}-I^\lambda_{B,n}
\end{array}\right)
\label{eq:Gij}
\ee
with $x_i-x_j$ for the argument of function $I$s.

\section{Equation for Single impurity bound state energies for zeroth and first angular momentum channels}

For a single impurity, we can drop the site index $i$ such that $\hat V^l_i=\hat V^l$ and $\overline{\psi_{i,l}}=\overline{\psi_{l}}$. When we assume that $\hat{V}^{l=-1,0,1}$ are the only non-zero components, we have following equations from the Eq. (4) of the main article.
\ba
\overline{\psi_0}&=&\overline{\hat{G}_0(E)}\hat{V}^0\overline{\psi_0}+\overline{\hat{G}_1(E)}\hat{V}^{-1}\overline{\psi_{-1}}+\overline{\hat{G}_{-1}(E)}\hat{V}^1\overline{\psi_1}\nonumber\\
\overline{\psi_1}&=&\overline{\hat{G}_0(E)}\hat{V}^1\overline{\psi_1}+\overline{\hat{G}_1(E)}\hat{V}^0\overline{\psi_0}\nonumber\\
\overline{\psi_{-1}}&=&\overline{\hat{G}_0(E)}\hat{V}^{-1}\overline{\psi_{-1}}+\overline{\hat{G}_{-1}(E)}\hat{V}^0\overline{\psi_0}\\
\overline{\psi_{2}}&=&\overline{\hat{G}_1(E)}\hat{V}^1\overline{\psi_1}\nonumber\\
\overline{\psi_{-2}}&=&\overline{\hat{G}_{-1}(E)}\hat{V}^{-1}\overline{\psi_{-1}}\nonumber
\ea
For a bound state solution to exist the following condition must be satisfied,
\be
\det\left[ \begin{array}{ccc}
\overline{\hat{G}_0(E)}\hat{V}^{-1}-1 & \overline{\hat{G}_{-1}(E)}\hat{V}^{0} & 0 \\
\overline{\hat{G}_1(E)}\hat{V}^{-1} & \overline{\hat{G}_{0}(E)}\hat{V}^{0}-1 & \overline{\hat{G}_{-1}(E)}\hat{V}^{1} \\
0 & \overline{\hat{G}_1(E)}\hat{V}^{0} & \overline{\hat{G}_0(E)}\hat{V}^{1}-1
\end{array}\right]=0.
\label{eq:s1}
\ee
One can solve the above equations in both analytic and numeric ways to get the bound state spectrum. Most of the results for the purely magnetic impurities in s and p-wave superconductors are provided in the main article.
Due to the broken spin-rotation symmetry, bound states of single magnetic impurity in p-wavw SC strongly depend on the direction of magnetic moment even in the absence of Rashba SOC. In the limit of zero SOC, one can find analytic solutions for the bound state spectrum.
\be
\frac{|E_{1,2}|}{\Delta_t}\!=\! \sqrt{\frac{2\!+\!2J_0J_1\cos^2\theta\!+\!J_0(J_0\!+\!J_0J_1^2\!\pm\!\sqrt{3J_1^2\!-\!2J_0J_1(1\!-\!J_1^2)\!+\!J_0^2(1\!+\!J_1^4)\!+\!J_1\cos 2\theta(J_1\cos 2\theta\!-\!2J_0(1\!+\!J_0\!-\!J_1^2))})}{2(1+J_0^2)(1+J_1^2)}}
\ee
Note that for $\alpha=0$ these bound states at are doubly degenerate.

\section{Effects of the scalar potential on the single impurity bound states in s-wave superconductor}
We now investigate the effect of the interplay between the scalar and
the magnetic potential of the impurity.
Without SOC, the effect of $U_{n}\neq 0$ is to merely shift the energy of the $l=n$ level\cite{balatsky2006}.
However, the presence of the SOC causes the scalar potential to qualitatively affect the spectrum
of the YSR states created by the magnetic potential ($J_0,J_1\neq 0$).
In the perturbative regime $\alpha \ll 1$, for ${\bf S}\parallel\hat z$,
we find that when $U_0\neq 0$ the energies of the $l=0,1$ states are given by the following analytical expressions:
%
%
\ba
\frac{|E_{l=0}|}{\Delta_s}&\approx &\frac{1\!-\!J_0^2\!+\!U_0^2}{\sqrt{(1\!-\!J_0^2\!+\!U_0^2)^2\!+\!4J_0^2}}\\
&+&\!\!\frac{4\tilde{\alpha}^2J_0^2J_1((1\!-\!J_0J_1)(1\!+\!J_0^2\!+\!U_0^2)\!+\!2J_0J_1U_0^2)}{((1\!-\!J_0J_1)(J_0\!+\!J_1)\!+\!J_1U_0^2)((1\!-\!J_0^2\!+\!U_0^2)^2\!+\!4J_0^2)^{\frac{3}{2}}}\nonumber\\
\frac{|E_{l=1}|}{\Delta_s}&\approx &\frac{1\!-\!J_1^2}{1\!+\!J_1^2}\!+\!\frac{4\tilde{\alpha}^2J_1^2(J_0(1\!-\!J_0J_1)\!+\!J_1U_0^2)}{(1\!+\!J_1^2)^2((1\!-\!J_0J_1)(J_0\!+\!J_1)\!+\!J_1U_0^2)}
\ea
whereas the energy of the $l=-1$ states remains unchanged. From these expressions we can see that the SOC correction to the energy of the $l=1$ level depends in a non-trivial way on $U_0$.
Analogously, we found that the energy of the $l=0$ level qualitatively depends on $U_1$.
To go beyond the perturbative regime we solved Eq.~\ref{eq:s1} with $U_0\neq 0$ numerically.
\begin{figure}[h]
\includegraphics[width=7cm]{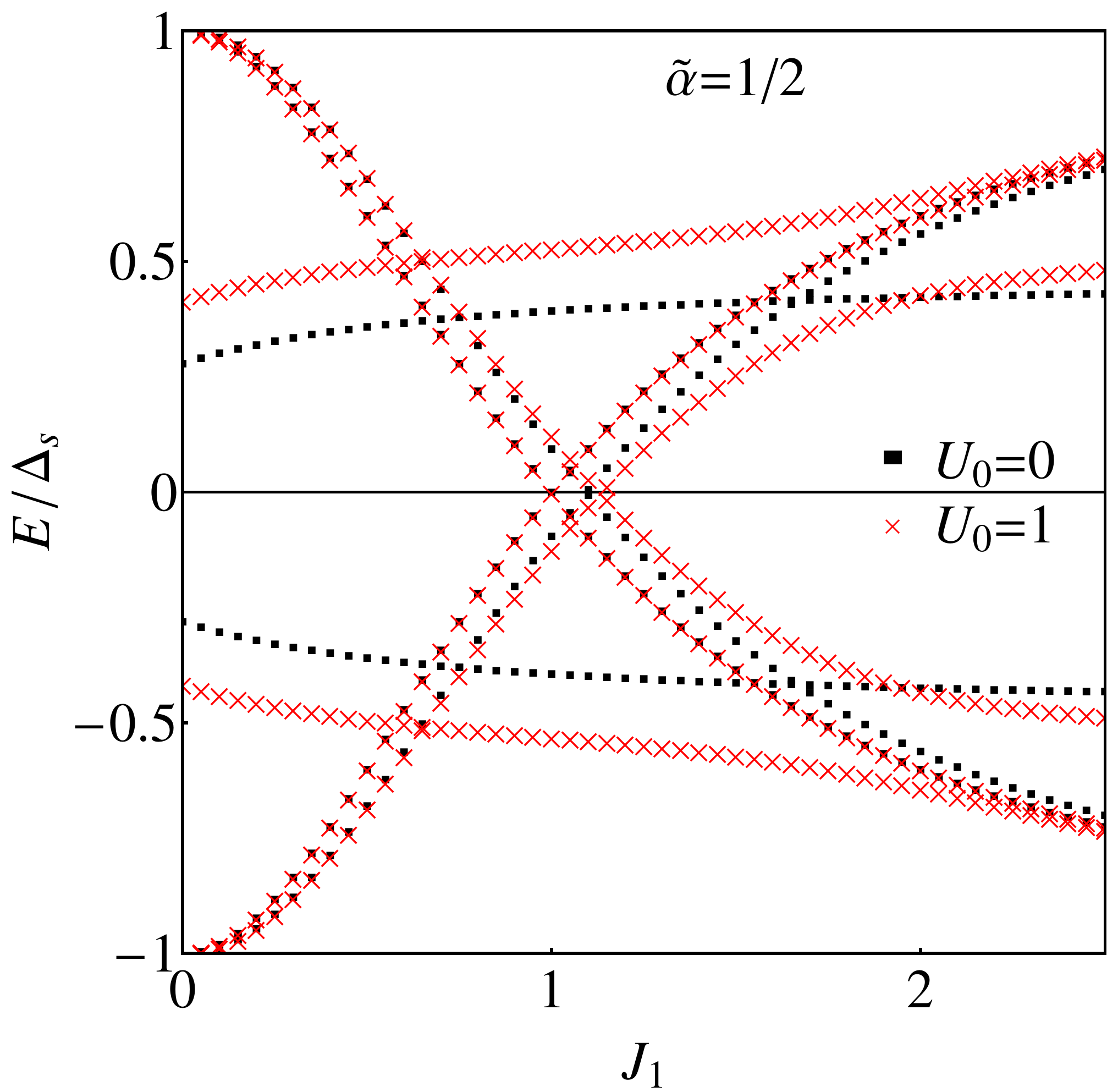}\,\,\,\,\,\,\,\,\,
\includegraphics[width=7cm]{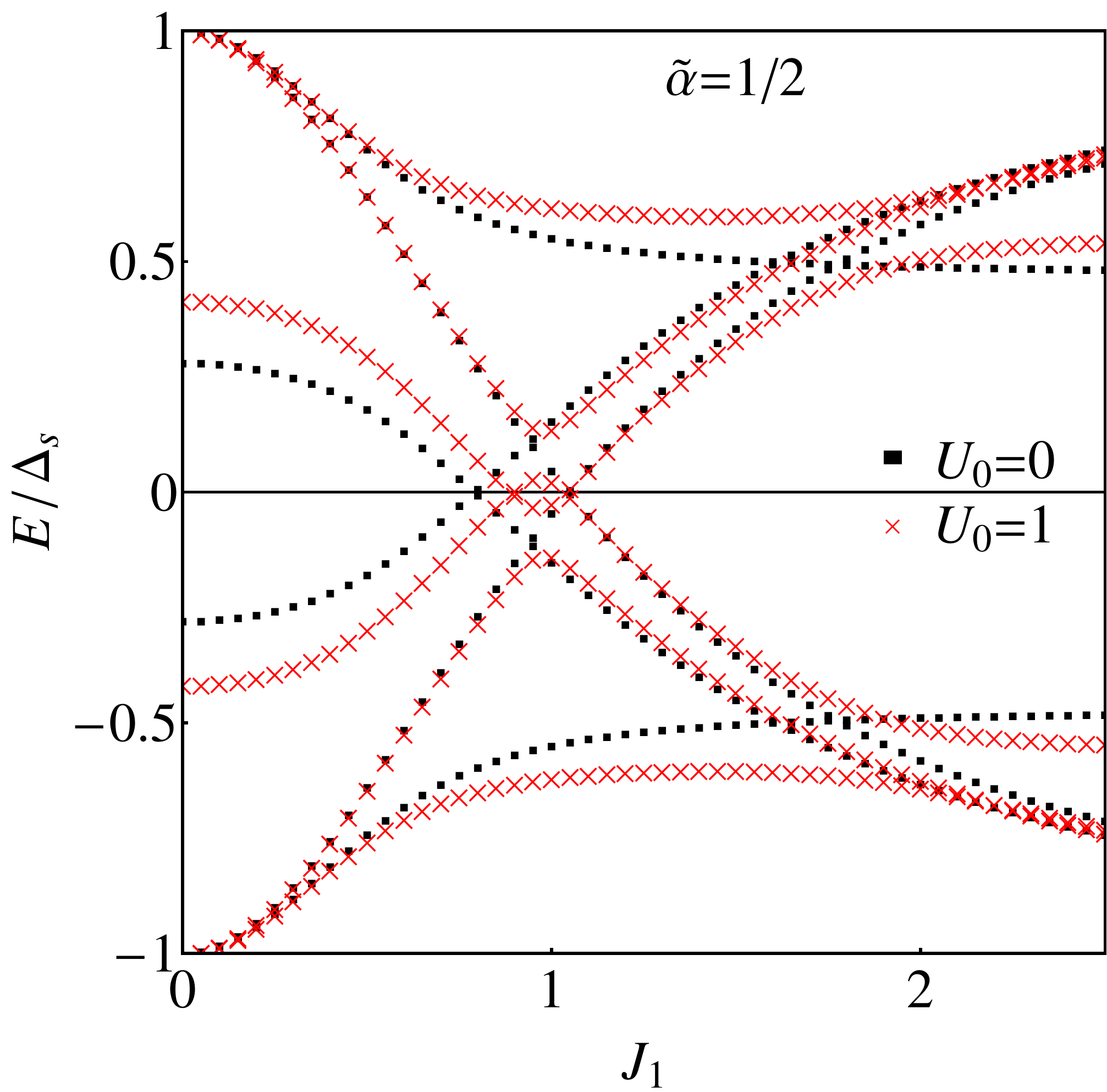}
\caption{(Color online) Bound state spectrum for a magnetic impurity with $U_0=0$ (black) and $U_0=1$ (red) in an s-wave SC as a function of $J_1$. $J_0=3/4$, $U_1=0$, $\vec{S}=\hat{z}$~(left panel) and $\vec{S}=\hat{x}$~(right panel) were used.}
\label{figs1}
\end{figure}
Fig.~\ref{figs1} shows the evolution of the YSR-states spectrum as a function of $J_1$
when both $U_0$ and $\tilde\alpha$ are not zero.
This figure clearly shows the qualitative effect that $U_0$ has on the YSR-spectrum
in the presence of SOC: for $\vec{S}\parallel\hat{z}$ the interplay of SOC and scalar potential
creates avoided crossings between the particle-like $l=0$ and the hole-like $l=1$ levels. For in-plane direction $\vec{S}$ there is an additional avoided crossing between particle-like and hole-like $l=1$ levels.

\section{Effects of the scalar potential on the single impurity bound states in p-wave superconductor}
\begin{figure}[h]
\includegraphics[width=7cm]{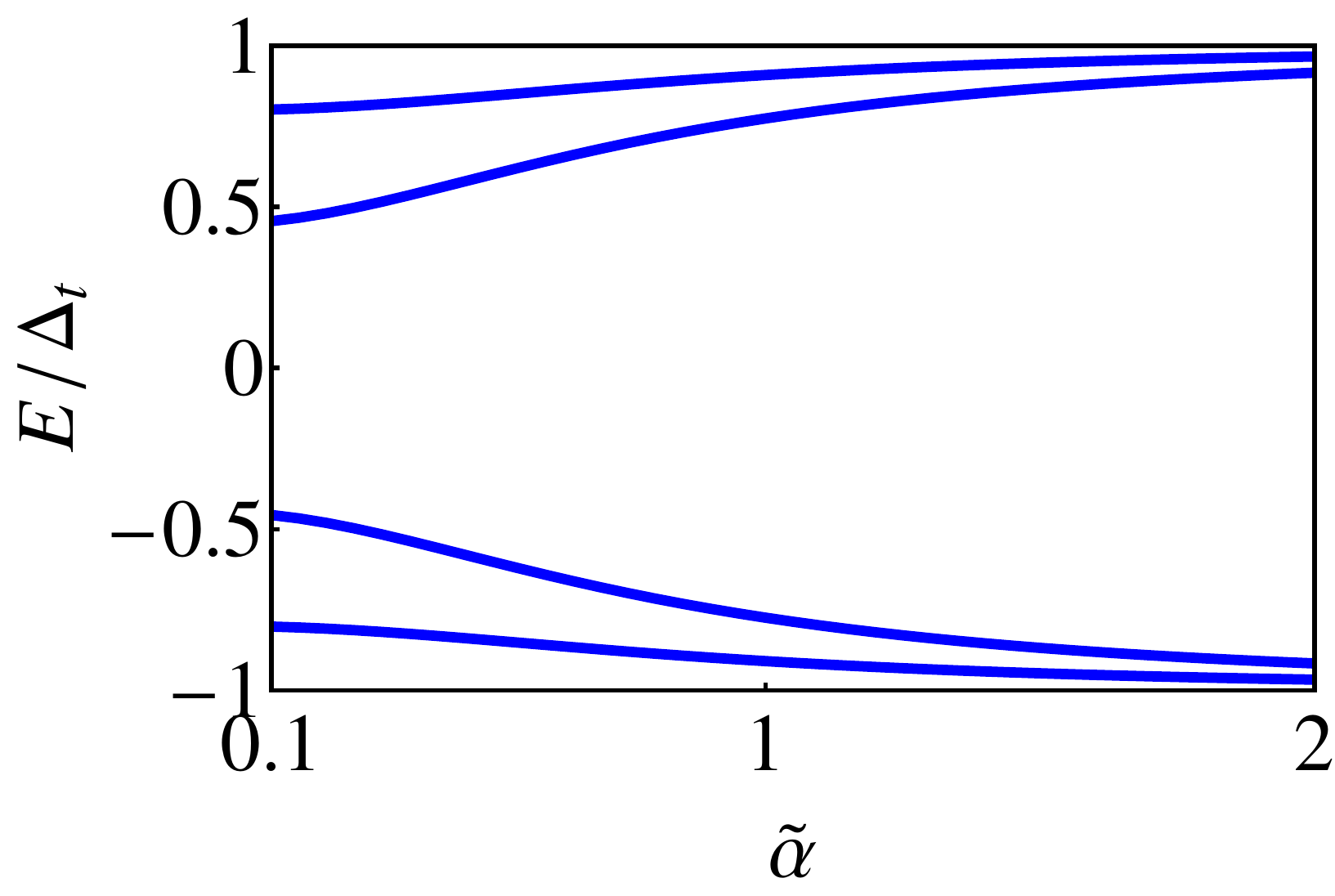}\,\,\,\,\,\,\,\,\,
\includegraphics[width=7cm]{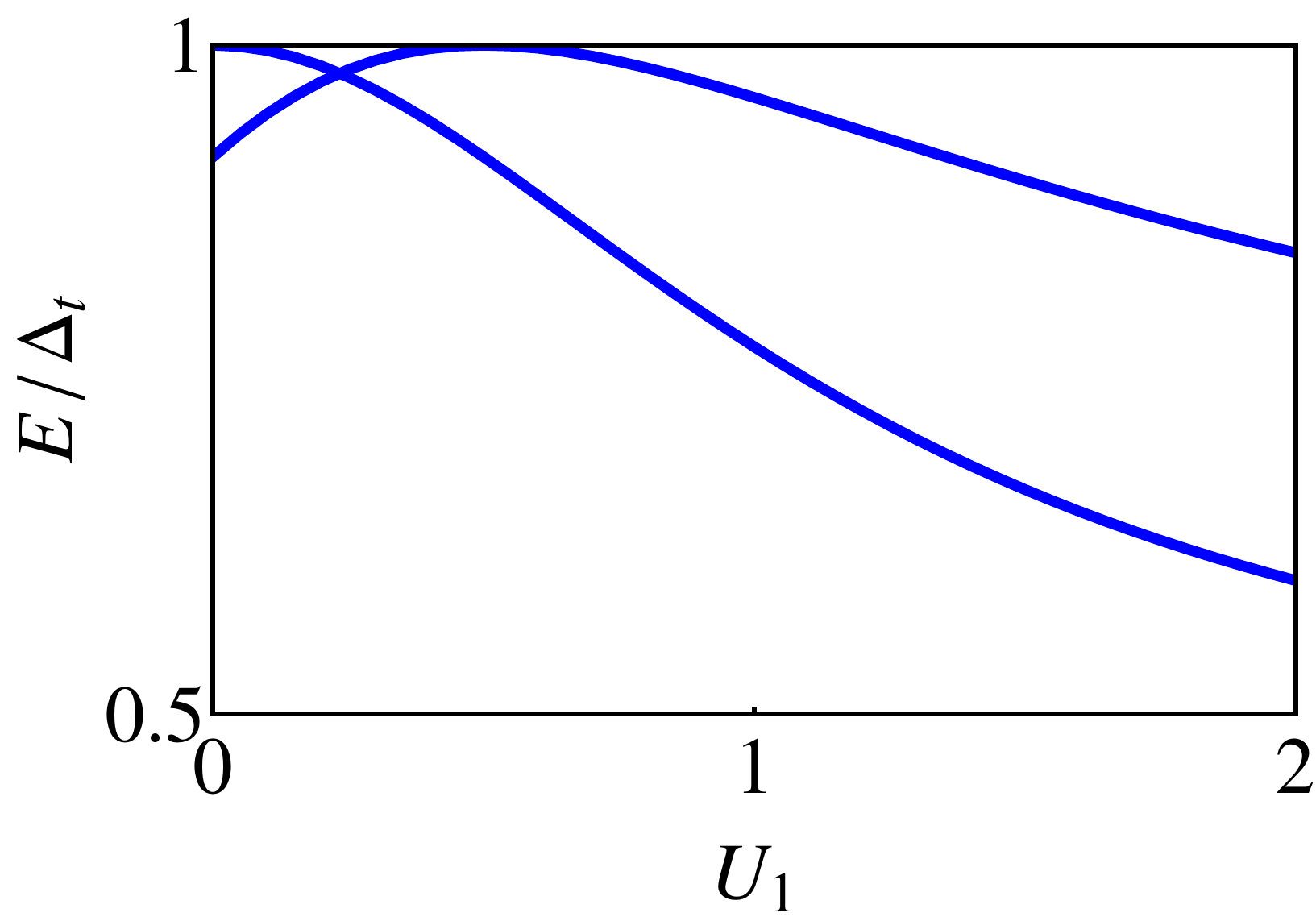}
\caption{(Color online) Bound state spectrum in a p-wave SC for a scalar impurity as a function of $\alpha$ at $U_0=0.5$ and $U_1=2$\,(left) and as a function of $U_1$ at $\alpha=0.5$ and $U_0=0.5$\,(right).}
\label{figs2}
\end{figure}

One can show that even in the presence of time-reversal symmetry, scattering off non-magnetic impurities alone leads to the formation of subgap bound states in the p-wave superconductors\cite{Wang04,Liu08,Nagai14}.
The presence of SOC modifies the spectrum. In the limit of no magnetic potential, for $\tilde\alpha \ll1$, we find the following analytical expressions for the energy levels of the bound states:

\begin{align}
\frac{|E_{l=0}|}{\Delta_t}&\approx\frac{U_0 U_1+1}{\sqrt{\left(U_0^2+1\right) \left(U_1^2+1\right)}} \label{eq:psc1}\\
&+\frac{\tilde\alpha^2 (U_0 - U_1)^2 ((U_0+U_1)^2+1 - U_0^2U_1^2)}{2 (1 +
   U_0 U_1) ((1 + U_0^2) (1 + U_1^2))^{3/2}}\nonumber\\
\frac{|E_{l=1}|}{\Delta_t}&\approx\frac{1+\tilde\alpha^2 U_1^2/2}{\sqrt{1 + U_1^2}}
 \label{eq:psc2}
\end{align}
Fig.~\ref{figs2}~(left) shows the evolution of these levels with $\tilde\alpha$.
In Fig.~\ref{figs2}~(right) we show the effect of $U_1$ for fixed values of $\tilde\alpha$ and $U_0$.
We see that there can be a value of $U_1$ for which the energy levels cross.
Notice that the bound state levels given by Eqs~\ref{eq:psc1},~\ref{eq:psc2} are doubly degenerate due to time reversal symmetry.
The presence of a magnetic potential ($J_n\neq 0$) leads to splitting of these Kramers doublets.

\section{Equation for Dimer bound state energies for zeroth and first angular momentum channels}
For dimer Eq. (4) of the main article becomes
\ba
\!\left(\!\begin{array}{ccc}
1-\overline{\hat G_0(E)}\hat V^{-1}_1 & -\overline{\hat G_{-1}(E)}\hat V^{0}_1 & 0 \\
-\overline{\hat G_1(E)}\hat V^{-1}_1 & 1-\overline{\hat G_0(E)}\hat V^{0}_1 & -\overline{\hat G_{-1}(E)}\hat V^{1}_1  \\
0 & -\overline{\hat G_1(E)}\hat V^{0}_1  & 1-\overline{\hat G_0(E)}\hat V^{1}_1
\end{array}\!\right)\!
\!\left(\!\begin{array}{c}
\overline{\psi_{1,-1}} \\
\overline{\psi_{1,0}} \\
\overline{\psi_{1,1}}
\end{array}\!\right)\!-\!\left(\!\begin{array}{ccc}
\overline{\hat G^{12}_{0}(E)}\hat V^{-1}_2 &\overline{\hat G^{12}_{-1}(E)}\hat V^{0}_2 & \overline{\hat G^{12}_{-2}(E)}\hat V^{1}_2 \\
\overline{\hat G^{12}_{1}(E)}\hat V^{-1}_2 & \overline{\hat G^{12}_{0}(E)}\hat V^{0}_2 & \overline{\hat G^{12}_{-1}(E)}\hat V^{1}_2 \\
\overline{\hat G^{12}_{2}(E)}\hat V^{-1}_2 & \overline{\hat G^{12}_{1}(E)}\hat V^{0}_2 & \overline{\hat G^{12}_{0}(E)}\hat V^{1}_2
\end{array}\!\right)\!\!\left(\!\begin{array}{c}
\overline{\psi_{2,-1}} \\
\overline{\psi_{2,0}} \\
\overline{\psi_{2,1}}
\end{array}\!\right)\!&=&0, \nonumber\\
-\!\left(\!\begin{array}{ccc}
\overline{\hat G^{21}_{0}(E)}\hat V^{-1}_1 &\overline{\hat G^{21}_{-1}(E)}\hat V^{0}_1 & \overline{\hat G^{21}_{-2}(E)}\hat V^{1}_1 \\
\overline{\hat G^{21}_{1}(E)}\hat V^{-1}_1 & \overline{\hat G^{21}_{0}(E)}\hat V^{0}_1 & \overline{\hat G^{21}_{-1}(E)}\hat V^{1}_1 \\
\overline{\hat G^{21}_{2}(E)}\hat V^{-1}_1 & \overline{\hat G^{21}_{1}(E)}\hat V^{0}_1 & \overline{\hat G^{21}_{0}(E)}\hat V^{1}_1
\end{array}\!\right)\!\!\left(\!\begin{array}{c}
\overline{\psi_{1,-1}} \\
\overline{\psi_{1,0}} \\
\overline{\psi_{1,1}}
\end{array}\!\right)\!+
\!\left(\!\begin{array}{ccc}
1-\overline{\hat G_0(E)}\hat V^{-1}_2 & -\overline{\hat G_{-1}(E)}\hat V^{0}_2 & 0 \\
-\overline{\hat G_1(E)}\hat V^{-1}_2 & 1-\overline{\hat G_0(E)}\hat V^{0}_2 & -\overline{\hat G_{-1}(E)}\hat V^{1}_2  \\
0 & -\overline{\hat G_1(E)}\hat V^{0}_2  & 1-\overline{\hat G_0(E)}\hat V^{1}_2
\end{array}\!\right)\!
\!\left(\!\begin{array}{c}
\overline{\psi_{2,-1}} \\
\overline{\psi_{2,0}} \\
\overline{\psi_{2,1}}
\end{array}\!\right)\!&=&0.\nonumber
\ea
Again bound-state energy is the solution of
\be
\det\left[ \begin{array}{cccccc}
\overline{\hat{G}_0(E)}\hat{V}^{-1}_1-1 & \overline{\hat{G}_{-1}(E)}\hat{V}^{0}_1 & 0 & \overline{\hat G^{12}_{0}(E)}\hat V^{-1}_2 & \overline{\hat G^{12}_{-1}(E)}\hat V^{0}_2 & \overline{\hat G^{12}_{-2}(E)}\hat V^{1}_2 \\
\overline{\hat{G}_1(E)}\hat{V}^{-1}_1 & \overline{\hat{G}_{0}(E)}\hat{V}^{0}_1-1 & \overline{\hat{G}_{-1}(E)}\hat{V}^{1}_1 & \overline{\hat G^{12}_{1}(E)}\hat V^{-1}_2 & \overline{\hat G^{12}_{0}(E)}\hat V^{0}_2 & \overline{\hat G^{12}_{-1}(E)}\hat V^{1}_2 \\
0 & \overline{\hat{G}_1(E)}\hat{V}^{0}_1 & \overline{\hat{G}_0(E)}\hat{V}^{1}_1-1 &\overline{\hat G^{12}_{2}(E)}\hat V^{-1}_2 & \overline{\hat G^{12}_{1}(E)}\hat V^{0}_2 & \overline{\hat G^{12}_{0}(E)}\hat V^{1}_2 \\
\overline{\hat G^{21}_{0}(E)}\hat V^{-1}_1 &\overline{\hat G^{21}_{-1}(E)}\hat V^{0}_1 & \overline{\hat G^{21}_{-2}(E)}\hat V^{1}_1 & \overline{\hat G_0(E)}\hat V^{-1}_2 -1 & \overline{\hat G_{-1}(E)}\hat V^{0}_2 & 0 \\
\overline{\hat G^{21}_{1}(E)}\hat V^{-1}_1 & \overline{\hat G^{21}_{0}(E)}\hat V^{0}_1 & \overline{\hat G^{21}_{-1}(E)}\hat V^{1}_1 & \overline{\hat G_1(E)}\hat V^{-1}_2 & \overline{\hat G_0(E)}\hat V^{0}_2 -1 & \overline{\hat G_{-1}(E)}\hat V^{1}_2  \\
\overline{\hat G^{21}_{2}(E)}\hat V^{-1}_1 & \overline{\hat G^{21}_{1}(E)}\hat V^{0}_1 & \overline{\hat G^{21}_{0}(E)}\hat V^{1}_1 & 0 & \overline{\hat G_1(E)}\hat V^{0}_2  & \overline{\hat G_0(E)}\hat V^{1}_2 -1
\end{array}\right]=0.
\ee
Due to the inter-site terms the spectrum now depends on the distance between two impurities as well as their directions. Please see the figure 3 in the main text.


\end{document}